\documentclass[lettersize,journal]{IEEEtran}
\usepackage{bm}
\usepackage[utf8]{inputenc}
\usepackage{graphicx}
\usepackage{epstopdf}
\usepackage{multirow}
\usepackage{amsmath}
\usepackage{tabularx}
\usepackage{booktabs}
\usepackage{float}
\usepackage[linesnumbered,algoruled,boxed,lined]{algorithm2e}
\usepackage{array}
\usepackage{color}
\usepackage{amssymb}
\usepackage{cite}
\usepackage{soul}

\usepackage{adjustbox}
\usepackage{setspace}
\usepackage{makecell}
\usepackage{bbding}
\usepackage{utfsym}
\usepackage{epstopdf}
\usepackage{tcolorbox}
\usepackage{ragged2e}
\usepackage{threeparttable}
\def\BibTeX{{\rm B\kern-.05em{\sc i\kern-.025em b}\kern-.08em
		T\kern-.1667em\lower.7ex\hbox{E}\kern-.125emX}}
\normalsize

\ifCLASSINFOpdf
\else
\fi
\begin{document}
	
	%
	\title{Diffusion-Model-enhanced Multiobjective Optimization for Improving Forest Monitoring Efficiency in UAV-enabled Internet-of-Things}
	%
	%
	%
	
\author{Hongyang Pan, Bin Lin,~\IEEEmembership{Senior Member,~IEEE}, Yanheng Liu, Shuang Liang,\\ and Chau Yuen,~\IEEEmembership{Fellow,~IEEE}
	
	\vspace{-1cm}
	\IEEEcompsocitemizethanks
	{
		\IEEEcompsocthanksitem This study is supported in part by the National Natural Science Foundation of China (62371085), in part by the Fundamental Research Funds for the Central Universities (3132023514), in part by the Science and Technology Development Plan Project of Jilin Province (20250102043JC), and in part by the National Research Foundation, Singapore and Infocomm Media Development Authority under its Future Communications Research \& Development Programme (FCP-NTU-RG-2024-025). (\textit{Corresponding authors: Bin Lin and Shuang Liang})
		\IEEEcompsocthanksitem Hongyang Pan and Bin Lin are with the Information Science and Technology College, Dalian Maritime University, Dalian 116026, China (E-mail: panhongyang18@foxmail.com; binlin@dlmu.edu.cn).
		\IEEEcompsocthanksitem Yanheng Liu is with the School of Computer Science, Zhuhai College of Science and Technology, Zhuhai 519000, China (E-mail: 16604747568@163.com).
		\IEEEcompsocthanksitem Shuang Liang is with the School of Information Science and Technology, Northeast Normal University, Changchun, 130117, China, and also with Key Laboratory of Symbolic Computation and Knowledge Engineering of Ministry of Education, Jilin University, Changchun 130012, China (E-mail: liangshuang@nenu.edu.cn).
		\IEEEcompsocthanksitem Chau Yuen is with the School of Electrical and Electronics Engineering, Nanyang Technological University, Singapore 639798 (E-mail: chau.yuen@ntu.edu.sg).
		\IEEEcompsocthanksitem This manuscript has been accepted by IEEE Internet of Things Journal, DOI: 10.1109/JIOT.2025.3590507.
	}
}
	%
	%

	\markboth{Journal of \LaTeX\ Class Files,~Vol.~14, No.~8, August~2015}%
	{Shell \MakeLowercase{\textit{et al.}}: Bare Demo of IEEEtran.cls for Computer Society Journals}
	%



	\maketitle
	
\begin{abstract}
	The Internet-of-Things (IoT) is widely applied for forest monitoring, since the sensor nodes (SNs) in IoT network are low-cost and have computing ability to process the monitoring data. To further improve the performance of forest monitoring, uncrewed aerial vehicles (UAVs) are employed as the data processors to enhance computing capability. However, efficient forest monitoring with limited energy budget and computing resource presents a significant challenge. For this purpose, this paper formulates a multi-objective optimization framework to simultaneously consider three optimization objectives, which are minimizing the maximum computing delay, minimizing the total motion energy consumption, and minimizing the maximum computing resource, corresponding to efficient forest monitoring, energy consumption reduction, and computing resource control, respectively. Due to the hybrid solution space that consists of continuous and discrete solutions, we propose a diffusion model-enhanced improved multi-objective grey wolf optimizer (IMOGWO) to solve the formulated framework. The simulation results show that the proposed IMOGWO outperforms other benchmarks for solving the formulated framework. Specifically, for a small-scale network with $6$ UAVs and $50$ SNs, compared to the suboptimal benchmark, IMOGWO reduces the motion energy consumption and the computing resource by $53.32\%$ and $9.83\%$, respectively, while maintaining computing delay at the same level. Similarly, for a large-scale network with $8$ UAVs and $100$ SNs, IMOGWO achieves reductions of $41.81\%$ in motion energy consumption and $7.93\%$ in computing resource, with the computing delay also remaining comparable.
\end{abstract}


%
%
\section{Introduction}
\par Forest monitoring is crucial for sustainable forest management \cite{DBLP:journals/lgrs/MarinelliCBB22}, and it can help conserve millions of hectares of forests \cite{sun2020learning}. In addition, it also benefits the lives of residents in nearby cities. For example, when wildfire strikes, the forest monitoring enables early detection, which can prevent loss of environment resources, property and lives. 

\par For a forest monitoring system, continuous monitoring is essential. Thus, the Internet-of-Things (IoT) is widely applied \cite{sun2020learning, DBLP:journals/ieeenl/LiaoSYAL20}. Specifically, an IoT network focuses on the use of sensor nodes (SNs), and aims to enhance efficiency, reduce costs, and optimize operations through the collection, analysis, and management of real-time data. Generally, an IoT network contains multiple SNs, such as temperature SNs, relative humidity SNs, and smoke SNs, where SNs can achieve local computing \cite{DBLP:journals/tii/ChooGP18}. By processing and analyzing vast amounts of data, IoT can provide actionable insights, which help make informed decisions and facilitate forest monitoring \cite{sun2020learning}. 

\par In addition, an effective and quick response is also vital for a forest monitoring system, since the system may be used to handle emergency incidents like wildfires \cite{DBLP:journals/tits/WangZW24}. To address the delay-sensitive issues, people have proposed mobile edge computing (MEC)-based solutions \cite{DBLP:journals/tpds/LiLXXJZZ22, DBLP:journals/tmc/MaLLJG22}. Specifically, many devices, such as satellites \cite{DBLP:journals/jsac/MaoLWGXWLK25, DBLP:journals/jsac/MaoLGXWLK25}, uncrewed aerial vehicles (UAVs) \cite{DBLP:journals/sj/MaoH021}, and ground mobile vehicles \cite{DBLP:journals/pieee/ZhangL20}, are exploited to assist MEC. Among them, using satellites as edge nodes is uneconomical, while using ground mobile vehicles will be limited by the topographical constraints of the forests. Thus, many researchers have employed UAVs for forest monitoring due to their low cost and strong line-of-sight communication capability \cite{sun2020learning, DBLP:journals/iotj/BushnaqCA21, DBLP:journals/taes/KaleemKAYAY24}. Furthermore, UAVs also hold great potential for wireless localization, which can further enhance their applicability in forest monitoring scenarios \cite{DBLP:journals/iotj/KatoRAS24}.  Additionally, compared to high-altitude long-range UAVs, low-altitude short-range UAVs can capture and update the first-hand information in real-time, assist SNs in computing, and feed back the computing results to IoT network \cite{DBLP:journals/comsur/PandeyGYJY25}. However, how to save the energy is a challenge due to the limited energy of UAVs.

\par Moreover, efficient resource management is crucial for UAV-enabled IoT in forest monitoring applications, where UAVs are deployed to collect, process, and transmit data from widespread and often inaccessible areas \cite{DBLP:journals/iotj/LiuYZGY24, DBLP:journals/iotm/MughalAAY24}. Under these circumstances, optimizing computing resource helps reduce the computational burden on UAVs \cite{niazmand2025joint, DBLP:journals/tvt/YeZZJSL18}, enabling the use of more lightweight and cost-effective platforms, which can reduce the overall system cost \cite{ye2021joint}. To simultaneously consider forest monitoring efficiency and system cost, these performance metrics should be comprehensively optimized. Thus, Pareto dominance can be introduced to provide a set of solutions, and then the decision-maker can balance the forest monitoring efficiency and system cost.
\par Since jointly considering these conflicting optimization objectives is difficult, the conventional algorithm will perform inefficiently. By integrating the diffusion model into the algorithm, it becomes possible to improve population diversity, and escape from local optima in complex optimization landscapes. Moreover, integrating the diffusion model is promising for the complex problems in UAV-enabled IoT scenarios, such as resource allocation \cite{tang2025dnn}, UAV deployment \cite{kawamoto2021uav, DBLP:journals/jsac/ZhaoLYLW24}, and task offloading decision \cite{DBLP:conf/wcnc/WuNTC0HX24}.

\par In this paper, we consider a UAV-enabled IoT system, in which multiple SNs are fixed on the ground to monitor the forests, and can perform local computation tasks. Owing to the limited computing capability of SNs, multiple UAVs can serve as aerial processors to provide a low-delay edge computing service. Compared to the conventional IoT system, UAV-enabled IoT system demonstrates high flexibility and scalability while leveraging the wide coverage of UAVs. However, the UAV-enabled IoT system for forest monitoring faces three challenges as follows. \textbf{First}, the system for monitoring forests requires a fast response, which means that the computing delay should be as small as possible. \textbf{Second}, the energy consumption of UAVs from the initial positions to the hovering positions should be reduced due to the limited on-board energy. \textbf{Finally}, the UAVs should occupy as little computing resource as possible to obtain greater economic benefits. To overcome the abovementioned challenges, we formulate a multi-objective optimization framework with three corresponding optimization objectives. The contributions of this paper are summarized as follows:

\begin{itemize}
	\item \emph{\textbf{UAV-enabled IoT system for forest monitoring and multi-objective optimization framework formulation}}: We consider a forest monitoring scenario, where multiple UAVs can provide ground SNs with a low-delay edge computing service. As the computing delay can influence the response time of forest monitoring, and the energy and resource are limited, we formulate a multi-objective optimization framework that aims to jointly minimize the maximum computing delay, minimize the total motion energy consumption, and minimize the maximum computing resource, while the constraints of total transmission power of SNs and UAV safe distance are considered.
	
	\item \emph{\textbf{A multi-objective heuristic algorithm for forest monitoring}}: As the formulated framework is a hybrid multi-objective optimization problem with both continuous and discrete solution spaces, we propose an improved multi-objective grey wolf optimizer (IMOGWO) with a diffusion model updating mechanism, a quasi-opposition based learning (QBL) strategy and a discrete solution updating mechanism to improve the performance of the algorithm. Specifically, the diffusion model updating mechanism enhances the quality of solutions in the archive, the QBL strategy improves the quality of solutions in the population, and the discrete solution updating mechanism is used to deal with the discrete solution.
	
	\item \emph{\textbf{Performance analysis}}: Through simulations, we can verify the effectiveness and performance of the proposed IMOGWO in both small- and large-scale networks. Specifically, for a small-scale network with $6$ UAVs and $50$ SNs, compared to the suboptimal benchmark, IMOGWO reduces the motion energy consumption and the computing resource by $52.32\%$ and $9.83\%$, respectively, while maintaining the computing delay at the same level. Similarly, for a large-scale network with $8$ UAVs and $100$ SNs, IMOGWO achieves reductions of $41.81\%$ in motion energy consumption and $7.93\%$ in computing resource, with the computing delay also remaining comparable.
\end{itemize}

\par The rest of this paper is organized as follows. The related work is reviewed in Section \ref{Related work}. Section \ref{System model} introduces the system model. Section \ref{Problem formulation} formulates a forest monitoring problem based on multi-objective optimization framework and solves it. Section \ref{Simulation} shows the simulation results. Finally, the conclusion is given in Section \ref{Conclusion}.

\section{Related Work}
\label{Related work}
\par In this section, we focus on the UAV forest communications, UAV-enabled IoT, and multi-objective optimization for UAV communications. In addition, the differences between previous works and this work are shown in Table \ref{Main_contributions_of_related_works}, and then the details are discussed as follows.

\begin{table*}[t]
	\setlength{\abovedisplayskip}{1pt}
	\setlength{\belowdisplayskip}{1pt}
	\setlength{\abovecaptionskip}{5pt}
	\setlength{\abovecaptionskip}{0.cm}
	\setlength{\belowcaptionskip}{-0.cm}
	\tabcolsep=0.01mm
	\tiny 
	\begin{center}
	\begin{threeparttable}
		\caption{Differences of previous works and this work}
		\begin{tabular}{|c|c|c|c|c|c|c|c|c|c|}
			\hline && \multicolumn{3}{c|}{$\begin{array}{c}\text{Optimization}\\\text {objectives}\end{array}$} &\multicolumn{4}{c|}{$\begin{array}{c}\text{Decision}\\\text {variables}\end{array}$} & Method 
			\\\hline &Reference & $\begin{array}{c}\text{Computing}\\\text {delay}\end{array}$ & $\begin{array}{c}\text{UAV} \\\text{motion}\\\text{energy}\\\text{consumption}\end{array}$ & $\begin{array}{c}\text{Computing}\\\text {resource}\end{array}$ & $\begin{array}{c}\text{UAV}\\\text {positions}\end{array}$ & $\begin{array}{c}\text{Resource}\\\text{or power}\\\text {allocation}\end{array}$  & $\begin{array}{c}\text{UAV-SN}\\\text {assignment}\end{array}$ &$\begin{array}{c}\text{Data}\\\text {split}\end{array}$ & $\begin{array}{c}\text{MOGWO,}\\\text {diffusion model}\\\text {updating mechanism,}\\\text{QBL strategy,}\\\text{and discrete}\\\text {solution updating} \\\text{mechanism}\end{array}$
			\\\hline  \multirow{7}{*}{$\begin{array}{c}\text{UAV forest}\\\text {communications}\end{array}$}&\cite{sun2020learning}  & $\checkmark$ & $\usym{2715}$ & $\usym{2715}$ &  $\usym{2715}$ & $\checkmark$ & $\checkmark$ & $\usym{2715}$ & $\usym{2715}$
			\\\cline{2-10}   &\cite{DBLP:journals/iotj/BushnaqCA21} &  $\checkmark$ & $\usym{2715}$  & $\usym{2715}$ & $\usym{2715}$  & $\usym{2715}$  & $\usym{2715}$ &$\usym{2715}$ & $\usym{2715}$ 
			\\\cline{2-10}  &\cite{DBLP:journals/access/ZhangDXCGD21}  & $\usym{2715}$ &  $\checkmark$ &  $\usym{2715}$ & $\checkmark$ &  $\usym{2715}$ &$\usym{2715}$ & $\usym{2715}$ &$\usym{2715}$ 
			\\\cline{2-10}  &\cite{WOS:000642319100001}  & $\checkmark$ & $\usym{2715}$ & $\usym{2715}$  &$\checkmark$ & $\usym{2715}$ & $\usym{2715}$ & $\usym{2715}$ & $\usym{2715}$ 
			\\\cline{2-10}  &\cite{chenxiao2022energy}  & $\usym{2715}$ & $\checkmark$ & $\usym{2715}$  &$\usym{2715}$ &  $\usym{2715}$ & $\usym{2715}$ &$\usym{2715}$ & $\usym{2715}$ 
			\\\cline{2-10}  &\cite{wang2023uav}  &$\usym{2715}$ & $\usym{2715}$ & $\usym{2715}$ & $\checkmark$ & $\usym{2715}$ & $\usym{2715}$ & $\usym{2715}$ &$\usym{2715}$
			\\\cline{2-10}  &\cite{DBLP:journals/tvt/BurhanuddinLDCZ22} &$\usym{2715}$ &$\usym{2715}$ &$\usym{2715}$ &$\checkmark$ &$\checkmark$ & $\usym{2715}$ &$\usym{2715}$ &$\usym{2715}$
			\\\hline  \multirow{11}{*}{$\begin{array}{c}\text{UAV-enabled}\\\text {IoT}\end{array}$}&\cite{DBLP:journals/sj/MaoH021}  & $\checkmark$ &$\usym{2715}$ &$\usym{2715}$ & $\checkmark$ & $\checkmark$ & $\checkmark$ & $\checkmark$ &$\usym{2715}$
			\\\cline{2-10}  &\cite{DBLP:journals/iotj/JeongC21} & $\usym{2715}$  & $\usym{2715}$ & $\usym{2715}$ &  $\checkmark$ &  $\checkmark$ & $\usym{2715}$ & $\usym{2715}$ & $\usym{2715}$
			\\\cline{2-10}  &\cite{DBLP:journals/tvt/LiuLLPD23}  & $\usym{2715}$ &  $\checkmark$ &  $\usym{2715}$ & $\checkmark$ & $\checkmark$ & $\checkmark$ &  $\usym{2715}$ &  $\usym{2715}$
			\\\cline{2-10}  &\cite{DBLP:journals/tvt/ChenSLC22}  &$\usym{2715}$ &$\usym{2715}$ &$\usym{2715}$ & $\checkmark$ & $\usym{2715}$ & $\checkmark$ &$\usym{2715}$&$\usym{2715}$
			\\\cline{2-10}  &\cite{DBLP:journals/tcom/XuZYLT21}  &$\usym{2715}$ &$\usym{2715}$ &$\usym{2715}$ &$\checkmark$ &$\checkmark$ &$\checkmark$ &  $\checkmark$ &$\usym{2715}$
			\\\cline{2-10}  &\cite{DBLP:journals/iotj/RosabalLPSHA22}  &$\usym{2715}$ &$\usym{2715}$ 	&$\usym{2715}$ 	& $\checkmark$ & $\checkmark$	&$\usym{2715}$&$\usym{2715}$ &$\usym{2715}$
			\\\cline{2-10}  &\cite{DBLP:journals/dcan/YuNZZ24}  &$\usym{2715}$ &$\usym{2715}$ &$\usym{2715}$ & $\checkmark$ &$\usym{2715}$ & $\checkmark$ & $\checkmark$ &$\usym{2715}$
			\\\cline{2-10}  &\cite{DBLP:journals/cn/WeiPXLYW24} & $\usym{2715}$ & $\usym{2715}$ & $\usym{2715}$  & $\checkmark$  & $\checkmark$ & $\usym{2715}$ & $\usym{2715}$ &$\usym{2715}$ 
			\\\cline{2-10}  &\cite{DBLP:journals/tgcn/GhdiriJAAY21}   & $\usym{2715}$  & $\checkmark$ & $\usym{2715}$  &  $\checkmark$ & $\usym{2715}$ &  $\checkmark$ & $\usym{2715}$&$\usym{2715}$
			\\\cline{2-10}  &\cite{DBLP:journals/iotj/HossainHA22}  & $\usym{2715}$ & $\usym{2715}$ & $\usym{2715}$ & $\usym{2715}$ & $\checkmark$ & $\checkmark$  & $\usym{2715}$ & $\usym{2715}$
			\\\hline  \multirow{6}{*}{$\begin{array}{c}\text{Multi-objective}\\\text{ optimization} \\\text{for UAV}\\\text{communications}\end{array}$} &\cite{chen2025joint}  & $\usym{2715}$ & $\checkmark$  & $\usym{2715}$ & $\checkmark$   & $\usym{2715}$ & $\checkmark$  & $\usym{2715}$& $\usym{2715}$
			\\\cline{2-10}  &\cite{DBLP:journals/tvt/HashirMMENK21}    &$\usym{2715}$ & $\usym{2715}$  & $\usym{2715}$ & $\checkmark$ & $\checkmark$ & $\usym{2715}$ & $\usym{2715}$  & $\usym{2715}$ 
			\\\cline{2-10}  &\cite{jian2024energy} & $\usym{2715}$ & $\usym{2715}$ & $\usym{2715}$ & $\checkmark$ & $\checkmark$ & $\checkmark$ & $\usym{2715}$ &$\usym{2715}$
			\\\cline{2-10}  &\cite{DBLP:journals/wcl/LiuLLLD23}  & $\usym{2715}$ & $\checkmark$  & $\usym{2715}$  & $\checkmark$ & $\checkmark$ & $\checkmark$& $\usym{2715}$  &$\usym{2715}$
			\\\cline{2-10} &\cite{DBLP:journals/tcom/ZhuZLLY24}  & $\usym{2715}$ & $\checkmark$ & $\usym{2715}$  & $\checkmark$ & $\usym{2715}$ & $\checkmark$ & $\usym{2715}$ & $\usym{2715}$ 
			\\\cline{2-10}  &\cite{DBLP:journals/wcl/PengHWK22}  & $\usym{2715}$ & $\checkmark$ & $\usym{2715}$ & $\checkmark$ & $\checkmark$ & $\usym{2715}$ & $\usym{2715}$& $\usym{2715}$
			\\\hline  &This work & $\checkmark$ &  $\checkmark$ & $\checkmark$ &  $\checkmark$ & $\checkmark$ &  $\checkmark$ &  $\checkmark$ & $\checkmark$ 
			\\\hline
		\end{tabular}
		\begin{tablenotes}
			\item The symbol ``$\checkmark$" signifies that the corresponding work explicitly formulates and optimizes the objective or decision variable considered in this work, while ``$\usym{2715}$" indicates that no such optimization is conducted in the corresponding work.
		\end{tablenotes}
		\vspace{-0.4cm}
		\label{Main_contributions_of_related_works}
	\end{threeparttable}
	\end{center}
\end{table*}

\subsection{UAV forest communications}

\par Sun \emph{et al}. \cite{sun2020learning} introduced a decomposition strategy including a novel particle swarm optimization algorithm and a Markov random field in their work to seek an optimal UAV resource allocation strategy. Bushnaq \emph{et al}. \cite{DBLP:journals/iotj/BushnaqCA21} investigated the effectiveness and reliability of the UAV-assisted IoT for forest fire detection. Specifically, the authors optimized IoT device density and UAV number to detect the forest fire to the greatest extent. The authors in \cite{DBLP:journals/access/ZhangDXCGD21} investigated the task planning for data collection by a UAV in forest monitoring. To minimize the UAV motion energy consumption, they proposed a bi-level hybrid meta-heuristic algorithm. The authors in \cite{WOS:000642319100001} analyzed the UAV trajectories for optimizing the network delay and communication throughput. For this purpose, an efficient architecture for the energy-efficient UAV communications was proposed, so as to monitor the forests. The authors in \cite{chenxiao2022energy} deployed UAVs as relay nodes between base station (BS) and users, and jointly optimized UAV relay transmission power and flight radius to achieve the highest energy efficiency under the forest environments. Wang \textit{et al.} \cite{wang2023uav} investigated the emergency UAV forest communications for the two cases, i.e., associated with a broken BS or an available BS. For this purpose, a hierarchical deep reinforcement learning (DRL) approach was proposed to optimize the UAV deployment and cluster the users. The authors in \cite{DBLP:journals/tvt/BurhanuddinLDCZ22} considered exploiting UAVs to capture videos of fire areas in the forest, and utilized UAV-to-UAV communications to transmit videos, where a DRL approach was proposed to improve the quality-of-experience.
\par However, the abovementioned works focused solely on a single optimization objective, whereas real-world scenarios often necessitate the simultaneous consideration of multiple competing optimization objectives, such as balancing computing delay, energy efficiency, and resource utilization, to achieve practical and robust system performance.

\subsection{UAV-enabled IoT}

\par Mao \emph{et al}. \cite{DBLP:journals/sj/MaoH021} investigated a satellite-UAV-enabled IoT system to reduce offloading delay. By employing block coordinate descent (BCD) and successive convex approximation, they ensured the convergence of their algorithm to effectively address the formulated problem. The authors in \cite{DBLP:journals/iotj/JeongC21} simultaneously sent data and energy to multiple SNs, in which the common data should be recovered by all SNs, while private data was recovered only by the dedicated SNs. The authors in \cite{DBLP:journals/tvt/LiuLLPD23} investigated a fair energy-efficient resource strategy for IoT network to optimize energy consumption of multiple UAVs. Chen \textit{et al.} \cite{DBLP:journals/tvt/ChenSLC22} studied a UAV-assisted data collection scenario, where they proposed a quality-of-service based power allocation algorithm to optimize the placement of a single UAV, while covering massive SNs. The authors in \cite{DBLP:journals/tcom/XuZYLT21} investigated the security problems for dual UAV-assisted MEC systems, where one UAV acted as a processor to compute the offloaded tasks and the other one served as a jammer to suppress the vicious eavesdroppers. For solving this problem, a BCD-based algorithm was proposed. The authors in \cite{DBLP:journals/iotj/RosabalLPSHA22} aimed at minimizing the average energy consumption under the worst case in a UAV-enabled IoT system, and then they proposed a joint optimization approach to obtain UAV deployment and the transmission power of SNs. Yu \textit{et al.} \cite{DBLP:journals/dcan/YuNZZ24} considered a UAV-assisted cooperative offloading energy efficiency system for MEC to minimize the user terminal energy consumption. Then, they decomposed the problem into three convex subproblems, and used BCD algorithm to solve them. The authors in \cite{DBLP:journals/cn/WeiPXLYW24} studied a UAV-enabled heterogeneous IoT system, and established a coverage maximization problem. For solving this, an improved particle swarm optimization algorithm was exploited. The authors in \cite{DBLP:journals/tgcn/GhdiriJAAY21} considered a multi-UAV-enabled IoT system, where UAVs collected data from time-constrained SNs. Then, they formulated a problem to minimize the deployment costs and energy consumption. The authors in \cite{DBLP:journals/iotj/HossainHA22} focused on the use of UAVs in large-scale IoT for MEC, and thus they introduced a dynamic numerology scheme assignment framework. Specifically, they formulated a multi-objective optimization problem to maximize the uplink spectral efficiency, while minimizing the energy consumption of SNs.

\par However, most of the abovementioned works also considered a single optimization objective. Even though some of abovementioned works, such as \cite{DBLP:journals/tgcn/GhdiriJAAY21, DBLP:journals/iotj/HossainHA22}, considered multiple optimization objectives, they only used the linear weighting method or the $\epsilon$-constraint method. Instead, we formulate the optimization framework based on Pareto dominance, which is practical but challenging. If the actual requirements change, such a formulation will obtain a solution set instead of a single solution, allowing the decision-maker to simply select a solution rather than re-run the algorithm, which enhances the portability of the scenario.

\subsection{Multi-objective optimization for UAV communications}
\par The authors in \cite{chen2025joint} investigated a UAV-assisted simultaneous wireless information and power transfer MEC system, where they optimized system energy efficiency and SN battery sustainability by a DRL approach. The authors in \cite{DBLP:journals/tvt/HashirMMENK21} aimed to maximize the achievable sum rate of all SNs, while minimizing the transmit power from UAV. Jian \emph{et al.} \cite{jian2024energy} aimed to realize energy efficient communication coverage enhancement for cooperative UAVs, and hence a multi-objective optimization algorithm was proposed to minimize the number and total energy consumption of UAVs while achieving load balance among UAVs. Liu \textit{et al.} \cite{DBLP:journals/wcl/LiuLLLD23} developed a UAV-enabled integrated sensing and communication system, where the energy efficiency and minimum radar mutual information were jointly considered. Zhu \emph{et al.} \cite{DBLP:journals/tcom/ZhuZLLY24} formulated a coverage utility and energy multi-objective optimization problem, and then proposed an enhanced multi-objective grey wolf optimizer (MOGWO) to solve it. Peng \emph{et al.} \cite{DBLP:journals/wcl/PengHWK22} introduced a constrained decomposition-based multi-objective evolutionary algorithm to address the joint energy-efficient offloading and safe path planning problem. To enhance the algorithm performance, they specifically leveraged infeasible individuals, which provided valuable information during the evolutionary process. 
\par However, none of these works considered jointly optimizing computing delay, UAV motion energy consumption, and computing resource. Such a consideration was more comprehensive for actual scenarios, as it addressed the diverse requirements of real-world applications. Specifically, minimizing computing delay ensures timely task results for time-sensitive applications (e.g., forest fire monitoring), while reducing motion energy consumption extends UAV operational duration, enhancing system endurance. Additionally, optimizing computing resource alleviates on-board computational burdens, enabling the use of cost-effective UAVs with lower processing capability. Thus, our goal is to develop a more efficient algorithm capable of minimizing computing delay, UAV motion energy consumption, and computing resource concurrently, while maintaining robustness across varying application scenarios.

\section{System Model and Preliminaries}
\label{System model}
\par In this section, the system overview, the forest transmission rate model, the local and edge computing delay model, the UAV energy consumption model, and multi-objective optimization problem are introduced. To enhance the readability, we list the important notations in Table \ref{table:notations}. 

\begin{table*}[htb]
	\setlength{\abovedisplayskip}{1pt}
	\setlength{\belowdisplayskip}{1pt}
	\setlength{\abovecaptionskip}{5pt}
	\setlength{\abovecaptionskip}{0.cm}
	\setlength{\belowcaptionskip}{-0.cm}
	\tiny
	\begin{center}
		\caption{List of important notations}
		\begin{tabular}{|c|c|c|c|}\hline
			Variable  &Physical meanings &Variable &Physical meanings\\\hline
			$M/\mathcal{M}$ & Number/Set of UAVs & $K/\mathcal{K}$ & Number/Set of SNs \\\hline
			$\mathcal{M}_{i}$ & $i$-th UAV & $\mathcal{K}_{j}$ & $j$-th SN \\\hline
			$D_{ij}$ & Distance between $\mathcal{M}_{i}$ and $\mathcal{K}_{j}$ & $D_{\text{forest}}$ & Total forest transmission distance \\\hline
			$D_{\text{free}}$ & Total free space transmission distance & ${PL}_{\text{total}}$ & Total path loss \\\hline
			${PL}_{\text{forest}}$ & Path loss of the forest & ${PL}_{\text{free}}$ & Path loss of the free space \\\hline
			${a}_{ij}$ & Assignment between $\mathcal{M}_{i}$ and $\mathcal{K}_{j}$ & $p_{j}$ & Transmission power of $\mathcal{K}_{j}$ \\\hline
			$R_{ij}$ & Transmission rate between $\mathcal{M}_{i}$ and $\mathcal{K}_{j}$ & $B$ & Bandwidth \\\hline
			${Ca}_{j}$ & Total size of computation task at $\mathcal{K}_j$ & $Ca_{\text{L}, j}$ & Size of computation task processed by local SN \\\hline
			$Ca_{\text{U}, j}$ & Size of computation task processed by UAV & $f_{\text{L}, j}$ & Local CPU-cycle frequency of $\mathcal{K}_j$ \\\hline
			$f_{\text{U}, ij}$ & Computing resource of $\mathcal{M}_i$ allocated to $\mathcal{K}_j$ & $T_{\text{L}, j}$  & Delay of processing $Ca_{\text{L}, j}$-bit task locally \\\hline
			$T_{\text{U}, j}$ & Delay for UAV edge processing & $P_\text{total}$ & Total transmission power limit of all SNs \\\hline
			$E_{i}$ & Motion energy consumption of $\mathcal{M}_{i}$ & $d_\text{s}$ & Safe distance for avoiding collision between two UAVs\\\hline
			${M_\text{L}}$ &Maximum computing delay of SNs & ${M_\text{U}}$ &Maximum computing delay of UAVs\\\hline
		\end{tabular}
		\vspace{-0.4cm}
		\label{table:notations}
	\end{center}
\end{table*}

\subsection{System overview}

\par
\begin{figure}[tbp]
	{\includegraphics[width=3.5in]{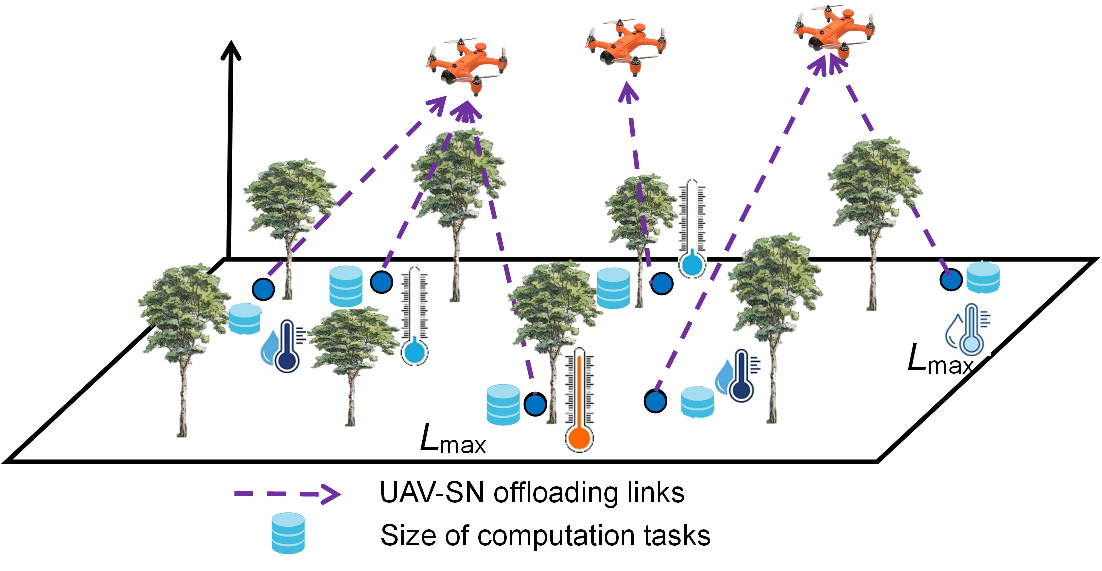}}
	\caption{Illustration of a UAV-enabled IoT system for monitoring forests.}
	\label{system_model}
\end{figure}

 To monitor the forests, multiple ground SNs that can collect data are deployed in an IoT network. As shown in Fig. \ref{system_model}, we consider a UAV-enabled IoT system, which consists of $M$ rotary-wing UAVs and $K$ SNs, denoted as $\mathcal{M}=\{1, 2, 3, ..., M\}$ and $\mathcal{K}=\{1, 2, 3, ..., K\}$, respectively. Specifically, SNs can execute some generated computation tasks when monitoring forests, while UAVs hover at their fixed positions to serve as processors to provide SNs with edge computing via wireless links \cite{8663615}. Without loss of generality, UAVs follow the fly-hover-communication protocol (FHCP) mentioned in \cite{DBLP:journals/iotj/LiuWSL22}. Moreover, when a UAV approaches its energy threshold, all UAVs are recalled, and a new batch is deployed. This batch-wise deployment strategy ensures seamless coverage and avoids service interruption due to energy shortages. Owing to their limited computing capability, SNs can offload computation tasks to the UAVs. The three-dimensional (3D) coordinates of the $i$-th UAV $\mathcal{M}_i$ and the $j$-th SN $\mathcal{K}_j$ are recorded as $\mathbb{Q}_i= (X_i, Y_i, Z_i)$ and $(x_j, y_j, 0)$, respectively. The distance between $\mathcal{M}_i$ and $\mathcal{K}_j$ can be expressed as follows: $D_{ij}=\sqrt{(X_i-x_j)^2+(Y_i-y_j)^2+Z_i^2}$. Besides, two UAVs should maintain a safe distance for avoiding collision, i.e., $\| \mathbb{Q}_i - \mathbb{Q}_{i^\prime} \| \geq d_{\text{s}}, \forall i \neq i^\prime $. 

\subsection{Forest transmission rate model}
\begin{figure}[htbp]
	\includegraphics[width=3.5in]{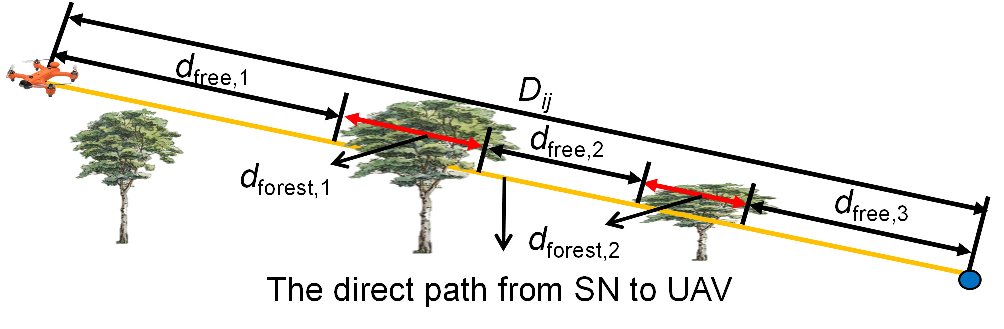}
	\caption{Modified model for forest environment.}
	\label{Modified_model_for_forest_environment}
\end{figure}
\par Referring to \cite{myagmardulam2021path}, we modify the empirical path loss model for the forest environment by combining it with the free space path loss model. As shown in Fig. \ref{Modified_model_for_forest_environment}, the distance between a UAV and an SN can be divided into two parts, which are the total forest transmission distance $D_{\text{forest}}$ and the total free space transmission distance $D_{\text{free}}$, respectively. Then, we have $D_{ij}=D_{\text{forest}}+D_{\text{free}}$. Specifically, these two parts can be computed as follows \cite{myagmardulam2021path}: $D_{\text{forest}}=\sum_{h=1}^{n}d_{\text{forest}, h}$ and $D_{\text{free}}=\sum_{u=1}^{m}d_{\text{free}, u}$, where $d_{\text{forest}, h}$ and $d_{\text{free}, u}$ are the $h$-th and $u$-th transmission distances in the forest and free space area, respectively. Then, the path loss of the forest and the free spaces can be expressed as follows \cite{myagmardulam2021path}:
\begin{subequations}
	\label{doublePL}
	\begin{align}
		{PL}_{\text{forest}}&=0.0021 f^{0.43}{D_{\text{forest}}}^{0.13},\\
		{PL}_{\text{free}}=-&27.56+20\lg(f)+20\lg(D_{\text{free}}),
	\end{align}
\end{subequations}
\noindent where $f$ is the frequency of the radio wave. Then, the total path loss is calculated as ${PL}_{\text{total}}={PL}_{\text{forest}}+{PL}_{\text{free}}$. Additionally, we assume that each SN can only offload its computation task to one UAV, while one UAV can assist multiple SNs \cite{DBLP:journals/twc/PanLSWY24}. For this purpose, we define $a_{ij} = 1$ to indicate that $\mathcal{K}_j$ decides to offload its computation task to $\mathcal{M}_i$, and $a_{ij}=0$ otherwise. Then, we have $\sum_{i=1}^M a_{ij}=1$. Moreover, we assume that the channels assigned to SNs served by different UAVs are orthogonal to each other, while SNs served by the same UAV are scheduled using time division multiple access (TDMA), which can further enhance the robustness of the communications \cite{10614110}. The transmission rate between $\mathcal{M}_i$ and $\mathcal{K}_j$ can be expressed as follows:
\begin{equation}
	\label{transmission rate}
	\begin{aligned}
		R_{ij}=a_{ij}B\log_2(1+\frac{p_j 10^{-{PL_{\text{total}, ij}}/10}}{\sigma^2}),
	\end{aligned} 
\end{equation}
\noindent where $B$ is the transmission bandwidth. $p_j$ is the transmission power of $\mathcal{K}_j$. $PL_{\text{total}, ij}$ is the total path loss between $\mathcal{M}_i$ and $\mathcal{K}_j$, and $\sigma^2$ is the power of the additive white Gaussian noise (AWGN). To extend the lifetime of the IoT network, we introduce a constraint $\sum_{j=1}^{K} p_j \leq P_{\text{total}}$ to limit the total transmission power of all SNs, where $P_{\text{total}}$ is the total transmission power budget of SNs. Moreover, if there are still SNs with depleted energy, UAVs can also be used for wireless charging \cite{DBLP:journals/ton/LinHYWWWZ23}.

\subsection{Local and edge computing delay model}
\par According to \cite{DBLP:journals/twc/BiZ18}, splitting the computation task does not result in additional input data. Thus, the entire $Ca_j$-bit computation task at $\mathcal{K}_j$ can be split into two parts $Ca_{\text{L}, j}$ and $Ca_{\text{U}, j}$, and we have $Ca_j=Ca_{\text{L}, j}+Ca_{\text{U}, j}$, where $Ca_{\text{L}, j}$ and $Ca_{\text{U}, j}$ are the size of computation tasks processed by local SN and UAV, respectively. Let $f_{\text{L}, j}$ and $f_{\text{U}, ij}$ denote the local central processing unit (CPU)-cycle frequency of $\mathcal{K}_j$ and the computing resource of $\mathcal{M}_i$ allocated to $\mathcal{K}_j$, respectively. Then, the delay of processing $Ca_{\text{L}, j}$-bit task locally is given as follows:
\begin{equation}
	\label{T^{l}_j}
	\begin{aligned}
		T_{\text{L}, j}= \frac{Ca_{\text{L}, j} {C}_j}{f_{\text{L}, j}},
	\end{aligned}
\end{equation}
\noindent where ${C}_j$ is the number of CPU cycles for computing $1$-bit data. Similarly, the delay for UAV edge processing consists of task offloading delay and task computing delay, which can be expressed as follows:
\begin{equation}
	\label{T^{u}_j}
	\begin{aligned}
		T_{\text{U}, j}= \sum_{i\in \mathcal{M}} a_{ij}\left(\frac{Ca_{\text{U}, j}}{R_{ij}}+ \frac{Ca_{\text{U}, j} {C}_j}{f_{\text{U}, ij}}\right).
	\end{aligned}
\end{equation}
\par Without loss of generality, the transmission delay from UAV to SN is ignored when UAV completes computing, since the transmission power of UAV is much larger than that of SN \cite{DBLP:journals/twc/BiZ18}.

\subsection{UAV energy consumption model}
\par The offloading energy consumption of UAVs is much smaller than the energy consumption to overcome the gravity and propulsion \cite{8663615}. Thus, in this paper, we only consider the energy consumption to overcome the gravity and propulsion, and the model proposed in \cite{8663615} for the two-dimensional (2D) horizontal space is given as follows:
\begin{equation}
	\label{UAV-2D-Power}
	\begin{split}
		P(V)=&P_{\text{B}}\left(1+\frac{3{V}^2}{U_{\text{tip}}^{2}}\right)+P_{\text{I}}\left(\sqrt{1+\frac{{V}^4}{4v_{0}^{4}}}-\frac{{V}^2}{2v_{0}^{4}}\right)^{\frac{1}{2}}\\+&\frac{1}{2}d_{0}\rho sAV^{3},
	\end{split}
\end{equation}
\noindent where $P_{\text{B}}$ and $P_{\text{I}}$ are two constant parameters about power in hovering status, respectively. $V$ refers to the velocity of the UAV from the initial position to the hovering position. Additionally, $U_{\text{tip}}$, $v_{0}$, $d_{0}$, $\rho$, $s$, and $A$ denote the tip speed of the rotor blade, the average rotor induction speed during hovering, the airframe drag ratio, the air density, the rotor solidity, and the area of the rotor disk, respectively. Note that we follow FHCP in this work, and hence it is reasonable to neglect the effects of UAV acceleration and deceleration, as they take only a negligible amount of time compared to the flight and hovering phases \cite{8663615, DBLP:journals/tmc/LiSDW24}.
\par Furthermore, this model is extended to 3D space, and the corresponding energy consumption model can be approximately expressed as follows \cite{8918497}: 
\begin{equation}
	\label{UAV-3D-Power}
	\begin{split}
		E(T)\approx &\int_{0}^{T} P(V(t))dt+\frac{M_{\text{UAV}}({V^2(T)}-{V^2(0)})}{2}\\+&M_{\text{UAV}}g(Z(T)-Z(0)),
	\end{split}
\end{equation}
\noindent where $V(t)$ represents the instantaneous UAV velocity at time $t$, and $T$ is the total flight time. $M_{\text{UAV}}$ and $g$ represent the mass and gravitational acceleration, respectively. Moreover, $Z(T)$ and $Z(0)$ represent the hovering and initial heights, respectively. 
\subsection{Multi-objective optimization problem}
\par In a multi-objective optimization problem, directly comparing the values of individual objectives can be challenging, as situations may arise where for two solutions $x_1$ and $x_2$, $f_a(x_1) < f_a(x_2)$ but $f_b(x_1) > f_b(x_2)$ 
\cite{DBLP:journals/tcom/SunLWWSL22}. To illustrate this more clearly, Fig. \ref{Example_diagram_BOP} shows an example of a bi-objective optimization problem. $x_4$ and $x_5$ are two solutions that satisfy the aforementioned relationship. In addition, we define that $x_5$ dominates $x_7$, denoted as $x_5 \preceq x_7$, if and only if $x_5$ is better than $x_7$ in at least one objective, and the mathematical expression for this dominance is given as follows:
\begin{figure}[htp]
	\centering
	\includegraphics[width=3in]{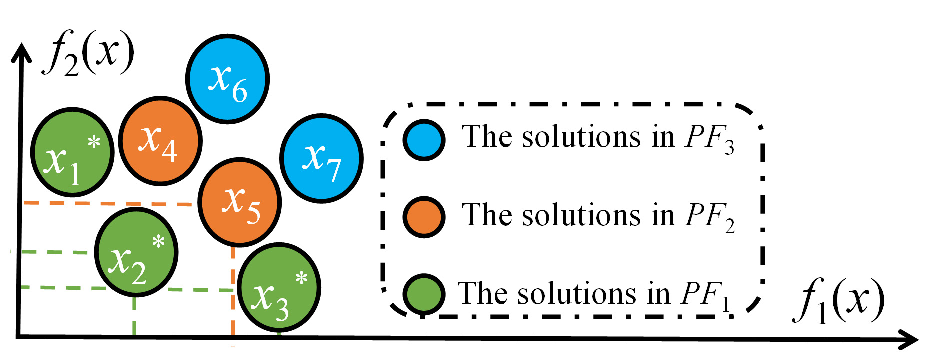}
	\caption{Example diagram of a bi-objective optimization.}
	\label{Example_diagram_BOP}
\end{figure}
\begin{subequations}
	\label{eqBOP}
	\begin{align}
		f_o(x_5)\leq f_o(x_7), \forall o\in {1, 2, 3,...,N_o},\\
		f_o(x_5)< f_o(x_7), \exists o\in {1, 2, 3,...,N_o},
	\end{align}
\end{subequations}
\noindent where $N_o$ is the number of objective functions. If there is no solution $x$ that satisfies $x \preceq x^*$, $x^*$ is a non-dominated solution. The set of non-dominated solutions includes all solutions that are not dominated by any other solutions. Typically, this set contains multiple non-dominated solutions rather than a single one, and the corresponding objective space of the non-dominated solution set is called the Pareto front (PF), i.e., $PF_1$ in Fig. \ref{Example_diagram_BOP}.

\section{Multi-objective Optimization Framework Formulation, Analysis and Solution}
\par In this section, the multi-objective optimization framework for forest monitoring is formulated, analyzed, and solved.
\label{Problem formulation}
\subsection{Framework formulation}
\par The optimization framework contains three optimization objectives. The first optimization objective is to minimize the maximum computing delay, which is decided by UAV positions, the transmission power of SNs, the computing resource allocation, the size of computation tasks processed locally, and the assignment between UAVs and SNs. However, changing UAV positions will cause additional motion energy consumption of UAVs, and hence the second optimization objective is to minimize the total motion energy consumption. Additionally, the third optimization objective is to minimize the maximum computing resource, so as to improve cost efficiency. Moreover, it also makes sense to jointly consider these decision variables. Specifically, the flexible 3D UAV deployment enhances communication quality and reduces transmission delay by enabling line-of-sight links. Meanwhile, the power allocation balances the IoT network lifetime and communication reliability, while task splitting and computing resource optimization reduce computing delay and utilize limited on-board resources efficiently. Besides, UAV-SN assignment can balance the loads across UAVs. The notations of the decision variables are detailed in Table \ref{table: Decision_Variables}.

\begin{table}[htbp]
	\begin{center}
		\caption{Notations of decision variables}
		\tiny
		\setlength{\tabcolsep}{0.1mm}
		{\begin{tabular}{|c|c|c|c|}\hline	
				\textbf{\begin{tabular}{c}Variables\end{tabular} } &\textbf{\begin{tabular}{c}Variable\\ elements\end{tabular}} & \textbf{\begin{tabular}{c}Physical\\ meanings\end{tabular}} & \textbf{Examples} \\\hline
				$\mathbb{Q}$ & $\{X_{i}, Y_{i}, Z_{i}| i \in \mathcal{M}\}$ & \begin{tabular}{c}Positions\\ of UAVs.\end{tabular} &\begin{tabular}{c} $(X_{1}, Y_{1}, Z_{1}) = (10, 20, 30)$ \\indicates that the position\\ of $\mathcal{M}_1$ is $(10, 20, 30)$.\end{tabular} \\\hline
				$\mathbb{P}$ & $\{p_{j}| j \in \mathcal{K}\}$ & \begin{tabular}{c}Transmission\\ power of SNs.\end{tabular}  &\begin{tabular}{c} $p_2 =0.3$ indicates that\\ the transmission power\\ of $\mathcal{K}_2$ is $0.3$ W.\end{tabular}\\\hline
				$\mathbb{F}$ &$\{f_{\text{U}, ij}|i \in \mathcal{M}, j \in \mathcal{K}\}$ & \begin{tabular}{c}Computing\\ resource \\ allocation.\end{tabular} &\begin{tabular}{c} $f_{\text{U}, 34} = 0.7$ indicates that\\ the computing resource of $\mathcal{M}_3$\\ allocated to $\mathcal{K}_4$ is $0.7$ GHz.\end{tabular}\\\hline
				$\mathbb{C}$ & $\{Ca_{\text{U}, j}| j \in \mathcal{K}\}$ & \begin{tabular}{c}Size of\\ computation\\ tasks processed\\ by UAVs. \end{tabular}& \begin{tabular}{c} $Ca_{\text{U}, 5} = 2^{20}$ indicates that\\ the size of computation\\ tasks of $\mathcal{K}_5$ processed\\ by the UAV is $2^{20}$ bits.\end{tabular}\\\hline
				$\mathbb{A}$ &$\{a_{ij}|i \in \mathcal{M}, j \in \mathcal{K}\}$ & \begin{tabular}{c} Assignment\\ between \\ UAVs and SNs. \end{tabular}&  \begin{tabular}{c}	$a_{67} = 1$ indicates that \\ $\mathcal{K}_6$ will offload its\\ computation tasks to $\mathcal{M}_7$.\end{tabular}
				\\\hline
		\end{tabular}}
		\label{table: Decision_Variables}
	\end{center}
\end{table}
\par \emph{\textbf{Optimization objective 1: minimizing the maximum computing delay.}} As mentioned above, the maximum computing delay is given by
\begin{equation}
	\label{f_1}
	\begin{aligned}
		f_1(\mathbb{Q}, \mathbb{P}, \mathbb{F}, \mathbb{C}, \mathbb{A}) = \max\{{M_\text{L}}, {M_\text{U}}\},
	\end{aligned}
\end{equation}
\noindent where ${M_\text{L}} = \max_{j\in \mathcal{K}} \{T_{\text{L}, j}\}$ is the maximum computing delay of SNs, and ${M_\text{U}} = \max_{i\in \mathcal{M}}\{ \sum_{j\in \mathcal{K}} T_{\text{U}, j} | a_{ij} =1\}$ is the maximum computing delay of UAVs.
\par \emph{\textbf{Optimization objective 2: minimizing the total motion energy consumption.}} The total motion energy consumption can be expressed as follows:
\begin{equation}
	\label{f_2}
	\begin{aligned}
		f_2(\mathbb{Q})=\sum_{i\in \mathcal{M}} \{E_{i}(T_{i})\},
	\end{aligned}
\end{equation}
\noindent where $T_{i}$ is the motion duration of $\mathcal{M}_i$ from the initial position to the hovering position, and $E_{i}$ represents the corresponding motion energy consumption.
\par \emph{\textbf{Optimization objective 3: minimizing the maximum computing resource.}} The computing resource of $\mathcal{M}_i$ is given by $f_{\text{U}, i}=\max_{j\in \mathcal{K}} \{a_{ij} f_{\text{U}, ij}\}$, and the third objective function can be described as follows:
\begin{equation}
	\label{f_3}
	\begin{aligned}
		f_3(\mathbb{F}, \mathbb{A})=\max_{i\in \mathcal{M}} \{f_{\text{U}, i}\}.
	\end{aligned}
\end{equation}
\par Thus, we formulate the framework as follows:
\begin{subequations}
	\label{Forest_monitoring_problem}
	\begin{align}
		\mathop{\text{min}}\limits_{\mathbb{Q}, \mathbb{P}, \mathbb{F}, \mathbb{C}, \mathbb{A}}&~f=\{f_1, f_2, f_3\}\\
		\text{s.t.} &\quad \mathcal{C}_1: L_{\min} \leq X_i \leq L_{\max}, \forall i\in \mathcal{M},\\
		\quad&\quad\mathcal{C}_2: L_{\min} \leq Y_i \leq L_{\max}, \forall i\in \mathcal{M},\\
		\quad&\quad\mathcal{C}_3: Z_{\min} \leq Z_i \leq Z_{\max}, \forall i\in \mathcal{M},\\
		\quad&\quad\mathcal{C}_4: p_{\min} \leq p_j \leq p_{\max}, \forall j\in \mathcal{K},\\
		\quad&\quad\mathcal{C}_5: f_{\text{U}, \min} \leq f_{\text{U}, ij} \leq f_{\text{U}, \max}, \forall j\in \mathcal{K},\\
		\quad&\quad\mathcal{C}_6: Ca_j=Ca_{\text{L}, j}+Ca_{\text{U}, j}, \forall j\in \mathcal{K},\\
		\quad&\quad\mathcal{C}_7: \sum_{i\in \mathcal{M}} a_{ij}=1, \forall j\in \mathcal{K},\\
		\quad&\quad\mathcal{C}_8: a_{ij}\in\{0, 1\}, \forall i\in \mathcal{M}, j\in \mathcal{K},\\
		\quad&\quad\mathcal{C}_9:  \sum_{j\in \mathcal{K}} p_j \leq P_{\text{total}}, \\
		\quad&\quad\mathcal{C}_{10}: ||\mathbb{Q}_i - \mathbb{Q}_{i^\prime}|| \geq d_{\text{s}}, \quad \forall i \neq i^\prime,
	\end{align}
\end{subequations}

\noindent where $\mathcal{C}_1$ and $\mathcal{C}_2$ limit the horizontal ranges of UAVs, in which $L_{\min}$ and $L_{\max}$ are the lower and upper bounds, respectively, while $\mathcal{C}_3$ limits their vertical ranges, in which $Z_{\min}$ and $Z_{\max}$ are the lower and upper bounds, respectively. $\mathcal{C}_4$ confines the transmission power of an SN, where $p_{\min}$ and $p_{\max}$ are the lower and upper bounds, respectively. Similarly, $\mathcal{C}_5$ limits the computing resource allocated to an SN by a UAV, where $f_{\text{U}, \min}$ and $f_{\text{U}, \max}$ are lower and upper bounds. Moreover, $\mathcal{C}_6$ is to ensure the entire data can be split, while $\mathcal{C}_7$ and $\mathcal{C}_8$ ensure that each SN is assigned to only one UAV. In addition, $\mathcal{C}_9$ is to limit the total transmission power of all SNs, and $\mathcal{C}_{10}$ is to avoid the collision between two UAVs.
\subsection{Framework analysis}
\par \textbf{Proposition 1.} \emph{(\ref{Forest_monitoring_problem}) is NP-hard, since the first optimization objective function $f_1$ can be simplified as a makespan minimization problem \cite{DBLP:journals/tase/SenguptaNS23}, which is NP-hard}.
\par \emph{\textbf{Proof:}} Suppose that no computation task needs to be processed locally by SNs, meaning that all computation tasks need to be transmitted to UAVs. In this case, given the fixed $\mathbb{Q}$, $\mathbb{P}$, $\mathbb{F}$, and $\mathbb{C}$, the first optimization objective can be transformed to a simplified optimization framework, and then the mathematical expression is as follows:
\begin{subequations}
	\label{S_Forest_monitoring_problem}
	\begin{align}
		\mathop{\text{min}}\limits_{\mathbb{A}}&~ \max\limits_{i\in \mathcal{M}}\{ \sum_{j\in \mathcal{K}} T_{\text{U}, j} | a_{ij} =1\}\\
		\text{s.t.} &\quad \mathcal{C}_7, \mathcal{C}_8.
	\end{align}
\end{subequations}
\noindent (\ref{S_Forest_monitoring_problem}) is a makespan minimization problem, which is NP-hard \cite{DBLP:journals/tase/SenguptaNS23}. Since (\ref{S_Forest_monitoring_problem}) is a simplified form with less optimization objectives and decision variables of (\ref{Forest_monitoring_problem}), (\ref{Forest_monitoring_problem}) is also NP-hard. $\hfill\blacksquare$

\par \textbf{Proposition 2.} \emph{(\ref{Forest_monitoring_problem}) is a mixed-integer non-linear programming problem, and hence it is non-convex.} 
\par \emph{\textbf{Proof:}} The whole solution of (\ref{Forest_monitoring_problem}) contains continuous solution $\mathbb{Q}, \mathbb{P}, \mathbb{F}$ and $\mathbb{C}$, and discrete solution $\mathbb{A}$. Thus, it is a mixed-integer non-linear programming problem (MINLP), which is non-convex. $\hfill\blacksquare$

\par \textbf{Remark 1.} \emph{The solution dimension of $\mathbb{F}$ and $\mathbb{A}$ can be reduced. Even though we reduce the dimension of $\mathbb{F}$ and $\mathbb{A}$, (\ref{Forest_monitoring_problem}) can still be a large-scale optimization problem \cite{DBLP:journals/tcyb/TianZZJ20}, since the solution dimension grows with the increasing number of UAVs or SNs.} 

\par
\begin{figure}[htbp]
	{\includegraphics[width=3in]{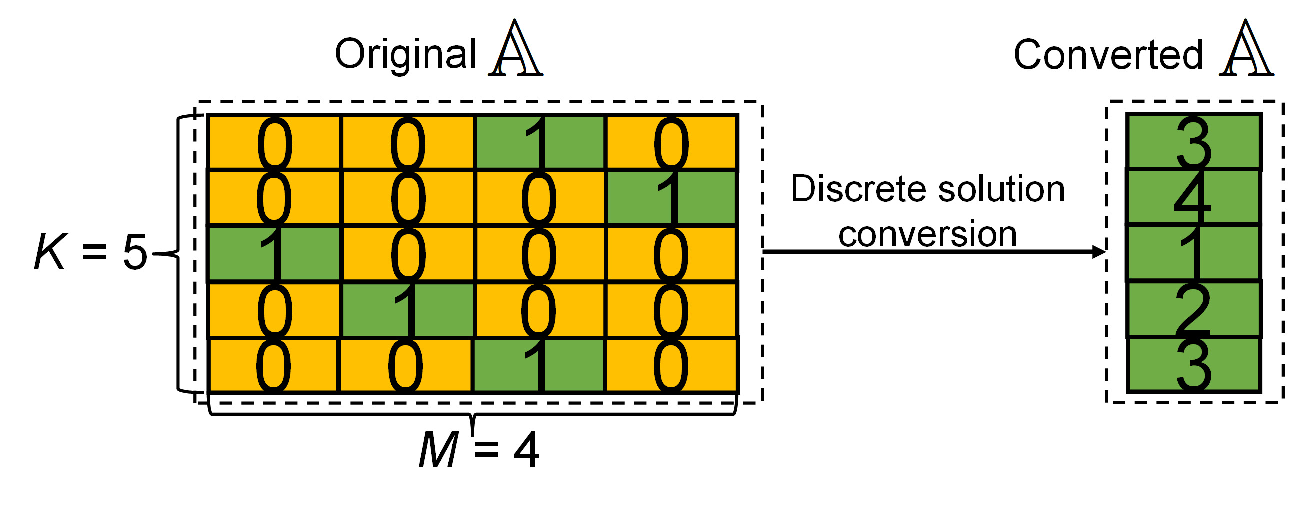}}
	\caption{An example of conversion from matrix to vector when $M=4$ and $K=5$.}
	\label{Example_matrix2vector}
\end{figure}

 \emph{\textbf{Analysis:}} The dimensions of $\mathbb{Q}, \mathbb{P}, \mathbb{F}, \mathbb{C}, \mathbb{A}$ are $3M$, $K$, $MK$, $K$, and $MK$, respectively. It is noted that we can convert the binary $\mathbb{A}$ into an integer vector. As shown in Fig. \ref{Example_matrix2vector}, we consider an example that contains $4$ UAVs and $5$ SNs, and thus the original $\mathbb{A}$ is a $5\times 4$ matrix. If the first SN $\mathcal{K}_1$ offloads its computation task to $\mathcal{M}_3$, which is shown in the first row of the matrix in Fig. \ref{Example_matrix2vector}, we record ``$3$" instead of the binary variables. Then, the solution dimension of the original $\mathbb{A}$ is reduced to $K$ due to the constraint that each SN will be assigned to a UAV. At this point, we complete the conversion between a matrix and a vector. Similarly, we can only allocate the computing resource for those specific $f_{\text{U}, ij}$ in which $i$ and $j$ satisfy $a_{ij} = 1$, and the rest of $f_{\text{U}, ij}$ are assigned to ``$0$''. Such a computing resource allocation is reasonable, since it will not increase the calculation results of Eq. (\ref{T^{u}_j}), and thus it will not increase $f_1$. Thus, the solution dimension of $\mathbb{F}$ can also be reduced to $K$, and the ultimate solution dimension of (\ref{Forest_monitoring_problem}) is $(3M+4K)$. Additionally, the optimization problem with more than $100$ decision variables is usually known as a large-scale optimization problem \cite{DBLP:journals/tcyb/TianZZJ20}. If we set $M=8$ and $K=100$, the solution dimension is $424$, indicating a large-scale optimization problem. $\hfill\blacksquare$

\par \textbf{Remark 2.} \emph{There are trade-offs between different optimization objectives.} 
\par \emph{\textbf{Analysis:}} On the one hand, minimizing the maximum computing delay requires UAVs with better communication links, which often necessitates that UAVs move to more suitable positions, thereby increasing the total motion energy consumption. On the other hand, if the computing resource of UAVs allocated to SNs is increased, the maximum computing delay will be reduced, while the maximum computing resource will be correspondingly increased. For ease of analysis, we only consider a simple case where all SNs will offload the data to UAVs, which means that $f_1$ can be transformed to maximize the transmission rate between UAVs and SNs, denoted as $f^\prime_1$. Moreover, other decision variables are given except UAV positions in the case, and we record the initial UAV positions and the optimal UAV positions for the maximum $f^\prime_1$ as $\mathbb{Q}^{\text{i}}$ and $\mathbb{Q}^{*}$, respectively. Since the decision-maker cannot predict the optimal solution of the problem in advance, the initial positions of UAVs are almost impossible to be the optimal positions, i.e., $\mathbb{Q}^{\text{i}}\neq \mathbb{Q}^{*}$. Obviously, if all UAVs remain at their initial positions, $f_2$ will reach its minimum, i.e., $f_2 = 0$. Instead, any attempt by UAVs to move to their optimal positions will cause $f_2 > 0$. Thus, increasing $f^\prime_1$, i.e. reducing $f_1$, will increase $f_2$. In addition, according to Eq. (\ref{T^{u}_j}), reducing $f_{\text{U}, ij}$ will reduce $f_3$, which will further increase $T_{\text{U}, j}$. Thus, $f_1$ will be correspondingly increased, which means that there is a trade-off between $f_1$ and $f_3$. $\hfill\blacksquare$

\subsection{IMOGWO for solving the framework}

\par In this section, the motivation to propose IMOGWO is introduced. Then, the details of the conventional MOGWO are presented. Afterward, IMOGWO is proposed. Finally, we utilize IMOGWO to solve (\ref{Forest_monitoring_problem}), and the algorithm is analyzed.
\subsubsection{Motivation}
\par There are three common methods to solve multi-objective optimization problems, which are DRL \cite{DBLP:journals/tccn/TangZXY24, DBLP:journals/iotj/FengWHY24}, convex optimization methods \cite{DBLP:journals/sj/MaoH021}, as well as multi-objective heuristic algorithms \cite{pan2025cooperative}, and different methods have their own suitable conditions for use. First, training a DRL model is time-consuming, which is appropriate to solve the problem with many continuous time slots \cite{DBLP:journals/tcom/PanLSFLY23}, \cite{DBLP:journals/twc/PanLSWY24}. However, (\ref{Forest_monitoring_problem}) is only with a single time slot, which means that it will take the unnecessary training time if we use DRL to solve it. In addition, convex optimization methods are not efficient to tackle an MINLP with multiple optimization objectives, such as (\ref{Forest_monitoring_problem}). Instead, multi-objective heuristic algorithms are good at solving large-scale optimization problems, which are non-convex and NP-hard. Specifically, multi-objective heuristic algorithms require significantly less time and cost for model training compared to DRL, and they can avoid reformulating the problem into a convex form so as to solve it with convex optimization methods, thereby preserving the original solution space. Moreover, the comparison method in multi-objective heuristic algorithms involves evaluating each optimization objective individually. As a result, it is unnecessary to standardize or unify the order of magnitude for the objectives.

\par Among many multi-objective heuristic algorithms, MOGWO \cite{mirjalili2016multi} is chosen as the basic skeleton to solve the formulated optimization framework, and the explicit reasons are as follows.
\begin{itemize}
	\item \textbf{\emph{Multi-leader hierarchy and adaptive position updating:}} MOGWO mimics the social hierarchy of grey wolves, where the top three wolves (alpha, beta, and delta) jointly guide the position updates of the rest of the population. Unlike many multi-objective heuristic algorithms that rely on a single elite solution to direct the search, the use of multiple leaders effectively diversifies guidance and reduces the risk of premature convergence. This strategy enhances the algorithm ability to explore complex landscapes and escape from local optima.
	
	\item \textbf{\emph{Effective balance between exploration and exploitation:}} MOGWO employs a dynamic encircling mechanism, wherein search agents adaptively shift between exploration and exploitation based on their proximity to the leaders. Such a balanced adaptability improves the global search capability of the algorithm, particularly in complex and high-dimensional optimization problems. The random nature of prey encirclement introduces flexibility, helping the algorithm avoid stagnation in the search space.
	
	\item \textbf{\emph{Simplicity and low parameter sensitivity:}} MOGWO has a simple structure with few control parameters, which eases its implementation and tuning across a wide range of optimization scenarios. Compared to algorithms with many control parameters, MOGWO requires fewer parameter fine-tuning efforts while maintaining competitive performance. This simplicity translates to faster execution time and lower computational overhead, which is crucial for real-time applications, especially in resource-limited environments such as UAV optimization tasks.
\end{itemize}
\noindent Then, the conventional MOGWO is briefly introduced.

\subsubsection{Conventional MOGWO}

\par MOGWO \cite{mirjalili2016multi} is a multi-objective version of grey wolf optimizer (GWO) \cite{DBLP:journals/aes/MirjaliliML14}, where GWO is inspired by grey wolves, which utilizes pursuing characteristics of grey wolves to find the best hunting position, and each solution is seen as a grey wolf. Specifically, MOGWO first chooses three leader grey wolves, recorded as $\vec{\chi}_\alpha$, $\vec{\chi}_\beta$, and $\vec{\chi}_\delta$. Then other grey wolves will learn from them. The explicit process can be mathematically described as follows:
\begin{subequations}
	\label{chase_the_prey}
	\begin{align}
		{G_{\alpha}}_{i, d}^{it}=|{O_1}_{i, d}^{it}\cdot {{\chi}_\alpha}_{i, d}^{it}-{\chi}_{i, d}^{it}|,\\
		{G_{\beta}}_{i, d}^{it}=|{O_2}_{i, d}^{it}\cdot {{\chi}_\beta}_{i, d}^{it}-{\chi}_{i, d}^{it}|,\\
		{G_{\delta}}_{i, d}^{it}=|{O_3}_{i, d}^{it}\cdot {{\chi}_\delta}_{i, d}^{it}-{\chi}_{i, d}^{it}|,\\
		{{\chi}_1}_{i, d}^{it}={{\chi}_\alpha}_{i, d}^{it}-{{U}_1}_{i, d}^{it}\cdot{G_{\alpha}}_{i, d}^{it},\\
		{{\chi}_2}_{i, d}^{it}={{\chi}_\beta}_{i, d}^{it}-{{U}_2}_{i, d}^{it}\cdot{G_{\beta}}_{i, d}^{it},\\
		{{\chi}_3}_{i, d}^{it}={{\chi}_\delta}_{i, d}^{it}-{{U}_3}_{i, d}^{it}\cdot{G_{\delta}}_{i, d}^{it},\\
		{\chi}_{i, d}^{it+1}=\frac{{{\chi}_1}_{i, d}^{it}+{{\chi}_2}_{i, d}^{it}+{{\chi}_3}_{i, d}^{it}}{3},
	\end{align}
\end{subequations}

\noindent where ${{\chi}_\alpha}_{i, d}^{it}$ means the element of $\vec{\chi}_\alpha$ for the $i$-th grey wolf in the $d$-th dimension at the $it$-th iteration. The similar expression is suitable for ${{\chi}_\beta}_{i, d}^{it}$ and ${{\chi}_\delta}_{i, d}^{it}$. ${U}$ (including $U_1, U_2, U_3$) and ${O}$ (including $O_1, O_2, O_3$) indicate the coefficients, which can be expressed as follows:
\begin{subequations}
	\label{coefficient_vectors}
	\begin{align}
		{U}=2{u}\cdot \text{rand}(1)-{u}\\
		{O}=2\cdot \text{rand}(1)
	\end{align}
\end{subequations}

\noindent where ${u}$ is a parameter, which decreases linearly from $2$ to $0$ following the iteration. $\text{rand}(1)$ is to generate a random number between $(0, 1)$. 
\par Although MOGWO has the abovementioned advantages, it is proposed to deal with the optimization problems only with continuous solution, which means that it cannot directly solve (\ref{Forest_monitoring_problem}). Moreover, the conventional multi-objective heuristic algorithms will fall into local optimality due to the complexity of the formulated framework. Thus, we exploit an IMOGWO to overcome the challenge and improve the quality of solution when solving the formulated optimization framework.

\subsubsection{IMOGWO}
\par In this section, an IMOGWO is proposed to solve (\ref{Forest_monitoring_problem}). Specifically, IMOGWO adopts three specific designs, which are the diffusion model updating mechanism, the QBL strategy, and the discrete solution updating mechanism. The explicit description of these specific designs is as follows, and the pseudocode of IMOGWO is shown in Algorithm \ref{IMOGWO}, where $G_{\max}$ as well as $Pop$ represent the maximum iteration and the population size, and $\mathcal{P}_{it}$ as well as $\mathcal{A}_{it}$ mean the population and the archive at the $it$-th iteration, respectively. 
\begin{algorithm}
	\caption{IMOGWO}
	\label{IMOGWO}
	\KwIn{$G_{\max}$, $Pop$, etc.}
	\KwOut{$\mathcal{A}_{G_{\max}}$.}
	$\mathcal{P}_{0}\Leftarrow \varnothing$, $\mathcal{A}_{0}\Leftarrow \varnothing$;\\
	Initialize $\mathcal{P}_{0}$ randomly;\\
	\While{$it\leq G_{\max}$}
	{	
		Choose the three leader grey wolves from $\mathcal{A}_{it}$;\\
		Update ($\mathbb{Q}, \mathbb{P}, \mathbb{F}, \mathbb{C}$) of $\mathcal{P}_{it}$ according to Eqs. (\ref{chase_the_prey}a)-(\ref{coefficient_vectors}b);\\
		Update ($\mathbb{A}$) of $\mathcal{P}_{it}$ according to Algorithm \ref{Discrete_solution_updating_mechanism}, and remain the better solutions in $\mathcal{P}_{it}$;\\		
		$\mathcal{Q}_{it}\Leftarrow \mathcal{P}_{it}$, update ($\mathbb{Q}, \mathbb{P}, \mathbb{C}$) of $\mathcal{Q}_{it}$ according to Eqs. (\ref{oppo}) and (\ref{oppo2}), and remain the better solution in $\mathcal{P}_{it}$;\\
		Update $\mathcal{A}_{it}$ based on $\mathcal{P}_{it}$;\\
		$\mathcal{Q}_{it}\Leftarrow \mathcal{A}_{it}$, update ($\mathbb{Q}, \mathbb{P}, \mathbb{C}$) of $\mathcal{Q}_{it}$ according to Eqs. (\ref{Ed_distance})-(\ref{newx}), and remain the better solutions in $\mathcal{A}_{it}$;\\
		Remain the non-dominated solutions in $\mathcal{A}_{it}$;\\
	}
	Return $\mathcal{A}_{G_{\max}}$.
\end{algorithm}

\par
\begin{figure}[tbp]
	\setlength{\abovedisplayskip}{1pt}
	\setlength{\belowdisplayskip}{1pt}
	\setlength{\abovecaptionskip}{1pt}
	\centering{\includegraphics[width=3.5in]{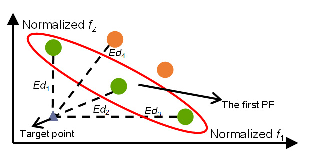}}
	\caption{An example to calculate Euclidean distance from each solution to target solution in a normalized bi-objective optimization problem.}
	\label{Ed}
\end{figure}

 \textit{(a) Diffusion model updating mechanism:} Diffusion model is suitable to update the archive during the iteration, and the reasons are as follows. First, the diffusion model can enrich the population diversity by introducing new candidate solutions into sparsely explored regions of the objective space, thereby improving the algorithm ability to escape local optima and maintain a well-distributed PF \cite{zhang2025diffusion}. Second, it is well-suited for high-dimensional tasks and can generate solutions that scale efficiently to larger problem sizes without architectural modifications. Finally, it is also lightweight and better suited for repeated use during iterative optimization. The details of using the diffusion model to update the archive are as follows. Specifically, we introduce an auxiliary variable $t = G_\text{max} - it +1$ to represent the reverse iteration process, and compute the probability density associated with each solution. Considering the three objectives, we first apply linear normalization to the objective values. Then, the Euclidean distance between each solution and the target point of the current iteration is computed as follows: 
\begin{equation}
	\label{Ed_distance}
	\begin{aligned}
		Ed^{t}_{i} = |\bar{F}^{t}_{i}- \bar{F}^{t}_{\text{min}}|,
	\end{aligned}
\end{equation}
\noindent where $\bar{F}^{t}_{i} = \{\bar{f_1}^{t}_{i}, \bar{f_2}^{t}_{i}, \bar{f_3}^{t}_{i}\}$ denotes the normalized objective vector of the $i$-th solution during the $(G_\text{max} - it +1)$-th iteration, and $\bar{F}_{t, \text{min}}$ corresponds to the normalized target point. An illustration of how the Euclidean distance is computed between each solution and the target point in a normalized bi-objective optimization is provided in Fig. \ref{Ed}. Following this, the probability density of each solution is estimated using an exponential function $PD^{t}_{i} = \operatorname{e}^{-Ed^{t}_{i}}/\sum\nolimits_{i=1}^{Pop} \operatorname{e}^{-Ed^{t}_{i}}$ \cite{DBLP:journals/icl/AlkheirI13}.
\par After calculating the probability density for each solution, the estimated $\hat{\chi}^{t}_{i}$ corresponding to the original solution $\vec{\chi}^{t}_{i} = \{\chi_{i, 1},... \chi_{i, dim}\}$ can be obtained, in which $dim$ is the dimension of the solution imposed by this mechanism:
\begin{equation}
\label{estimated_value}
\begin{aligned}
	\hat{\chi}^{t}_{i} = \frac{1}{Z} \sum\nolimits_{j=1}^{Pop} PD^{t}_{i} \mathcal{CN}(\vec{\chi}^{t}_{i}; \sqrt{\alpha^t}\vec{\chi}^{t}_{j}, 1-\alpha^t) \vec{\chi}^{t}_{j},
\end{aligned}
\end{equation}
\noindent  where $Z = \sum\nolimits_{j=1}^{Pop} PD^{t}_{i} \mathcal{CN}(\vec{\chi}^{t}_{i}; \sqrt{\alpha^t}\vec{\chi}^{t}_{j}, 1-\alpha^t)$ is the normalization term. The Gaussian kernel $\mathcal{CN}(\vec{\chi}^{t}_{i}; \sqrt{\alpha^t}\vec{\chi}^{t}_{j}, 1-\alpha^t)$ incorporates locality by weighting each solution $\vec{\chi}^{t}_{j}$ according to its distance from $\vec{\chi}^{t}_{i}$ in the parameter space \cite{zhang2025diffusion}. $\mathcal{CN}(\vec{\chi}^{t}_{i}; \sqrt{\alpha^t}\vec{\chi}^{t}_{j}, 1-\alpha^t)$ is a Gaussian kernel that introduces locality, effectively weighting the contribution of each solution $\vec{\chi}^{t}_{j}$ based on its proximity to $\vec{\chi}^{t}_{i}$ in the parameter space \cite{zhang2025diffusion}. The parameter $\alpha^t= \cos^2 \left({\pi t}/{2G_\text{max}}\right)$ gradually increases from $0$ to $1$ as $t$ decreases. Using $\hat\vec{\chi}^{t}_{i}$, we can generate new solutions through the diffusion model \cite{DBLP:conf/iclr/SongME21}: 
\begin{equation}
	\label{newx}
	\begin{aligned}
		\vec{\chi}^{t-1}_{i} = \sqrt{\alpha^{t-1}} \hat{\chi}^{t}_{i} + \sqrt{1- \alpha^{t-1} - {\sigma^t}^2} \frac{\vec{\chi}^{t}_{i}- \sqrt{\alpha^{t}}\hat{\chi}^{t}_{i}}{\sqrt{1- \alpha^{t}}} + \sigma^t \text{w}, 
	\end{aligned}
\end{equation}
\noindent where $\sigma^t \text{w}$ is a random mutation for the solution, and $\sigma^t = \sqrt{ \max\left( 0,\ \left( \frac{1 - \alpha^{t+1}}{1 - \alpha^t} - 1 \right)\right)  \cdot (1 - \alpha^{t+1}) }$ as well as $\text{w} \sim \mathcal{CN}(0, 1)$.
\par \textit{(b) QBL strategy:} To further enhance the search capability of the algorithm, we introduce a QBL strategy to update the continuous solution, which can exploit the search space of the continuous solution \cite{zheng2024reliable}. The details are as follows. 
\par \emph{\textbf{First}}, the opposite decision variable ${\chi^{\prime}}_{i, d}$ of the original decision variable ${{\chi}}_{i, d}$ is generated as follows: 
\begin{equation}
	\label{oppo}
	\begin{aligned}
		{\chi^{\prime}}_{i, d}=UB_{i, d}+LB_{i, d}-{{\chi}}_{i, d},\\
	\end{aligned}
\end{equation}
\noindent where $UB_{i, d}$ and $LB_{i, d}$ are the upper and lower bounds of the $i$-th solution in the $d$-th dimension. For example, $UB_{i, d}$ and $LB_{i, d}$ will be assigned $p_{\max}$ and $p_{\min}$ if we use QBL strategy to update $\mathbb{P}$.
\par \emph{\textbf{Second}}, the strategy gains the quasi-opposite decision variable ${\chi^{\prime\prime}}_{i, d}$, which can be expressed as follows: 
\begin{equation}
	\label{oppo2}
	\begin{aligned}
		{\chi^{\prime\prime}}_{i, d}={Mid}_{i, d}+ \text{rand}(1)\cdot ({\chi^{\prime}}_{i, d}-{Mid}_{i, d}),\\
	\end{aligned}
\end{equation}
\noindent where ${Mid}_{i, d}$ is the middle point, and ${Mid}_{i, d} = LB_{i, d}+(UB_{i, d}-LB_{i, d})/2$. Since QBL strategy does not contain manually adjusted parameters, it is parameter-insensitive.

\par \textit{(c) Discrete solution updating mechanism:} Each solution contains discrete $\mathbb{A}$, which cannot be tackled by the conventional MOGWO. Therefore, a discrete solution updating mechanism is proposed. The pseudocode is shown in Algorithm \ref{Discrete_solution_updating_mechanism}, where $\mathcal{DP}_{i}$ is the $i$-th discrete solution of the population. Specifically, if $\operatorname{rand}< \sigma_1$, the discrete solution will remain unchanged. If $\sigma_1 \leq\operatorname{rand}<\sigma_2$, the discrete solution of the archive in the current iteration will be assigned, which means that it will not generate the infeasible solutions. If $\operatorname{rand} \geq\sigma_2$, the discrete solution will be regenerated randomly.
\begin{algorithm}[tb]
	\caption{Discrete solution updating mechanism}
	\label{Discrete_solution_updating_mechanism}
	\KwIn{$\mathcal{DP}_{i}$, $\mathcal{A}_{it}$, $\sigma_1$, $\sigma_2$.}
	\KwOut{$\mathcal{DP}_{i}$.}
	\uIf{$\operatorname{rand}<\sigma_1$}{
		$\mathcal{DP}_{i} = \mathcal{DP}_{i}$;\\		
	}
	\uElseIf{$\sigma_1 \leq\operatorname{rand}(1)<\sigma_2$}{
		Randomly select a solution from $\mathcal{A}_{it}$, and record its discrete solution as $\mathcal{DP}^{\prime}_{i}$;\\
		$\mathcal{DP}_{i} = \mathcal{DP}^{\prime}_{i}$;\\	
	}
	\Else{
		Regenerate $\mathcal{DP}_{i}$ randomly;\\
		
	}
	Return $\mathcal{DP}_{i}$.
\end{algorithm}

\par \textbf{Remark 3.} \emph{The algorithm is able to tackle all constraints.}
\par \textbf{\emph{Analysis:}} $\mathcal{C}_1-\mathcal{C}_8$ are strong constraints that limit the maximum and minimum bounds or the domains of the decision variables. Among them, $\mathcal{C}_6$ can be easily handled when we generate a new solution. When $Ca_{\text{U}, j}$ is generated, we calculate $Ca_{\text{L}, j}$ by $Ca_{\text{L}, j}=Ca_j-Ca_{\text{U}, j}$, and thus $\mathcal{C}_6$ is satisfied. Through the conversion from the binary variables to the integer variables in \textbf{Remark 1}, $\mathcal{C}_7$ and $\mathcal{C}_{8}$ are satisfied. Moreover, $\mathcal{C}_{9}$ and $\mathcal{C}_{10}$ are weak constraints that will be regarded as a penalty function. Once $\mathcal{C}_{9}$ or $\mathcal{C}_{10}$ is not satisfied, the values of the three optimization objectives will be multiplied by a penalty constant $C^{\prime}>1$, i.e., $f_1 = C^{\prime}f_1$, $f_2 = C^{\prime}f_2$, and $f_3 = C^{\prime}f_3$. Since the values of these three optimization objectives are non-negative, the algorithm discards these solutions with poor quality during the iteration process. $\hfill\blacksquare$

\subsubsection{Solving the framework with IMOGWO}
\par Following the FHCP and the batch-wise UAV deployment strategy, IMOGWO is executed in a centralized manner and deployed at a central controller. The central controller can collect the global information, such as the SN positions and the total size of computation tasks, from the UAVs when the first batch of UAVs is deployed. The details of solving (\ref{Forest_monitoring_problem}) with IMOGWO are as follows:
\begin{itemize}
	\item \textbf{\textit{Step 1:}} Initialize the proposed IMOGWO randomly, and update the archive.
	\item \textbf{\textit{Step 2:}} Choose the three leader solutions from the archive. Then, update the continuous solution ($\mathbb{Q}, \mathbb{P}, \mathbb{F}, \mathbb{C}$) of the population according to the conventional MOGWO, and remain better solutions. 
	\item \textbf{\textit{Step 3:}} Update the discrete solution ($\mathbb{A}$) of the population according to the discrete solution updating mechanism, and remain better solutions. 
	\item \textbf{\textit{Step 4:}}	Update the continuous solution ($\mathbb{Q}, \mathbb{P}, \mathbb{C}$) of the population according to QBL strategy, remain better solutions, and update the archive. 
	\item \textbf{\textit{Step 5:}} Update the continuous solution ($\mathbb{Q}, \mathbb{P}, \mathbb{C}$) of the archive according to the diffusion model updating mechanism, remain the better solutions, and update the archive according to the non-dominated sorting. 
	\item \textbf{\textit{Step 6:}} If the termination condition is not met, IMOGWO will repeat the \textbf{\textit{Steps 2-5}}. Otherwise, the solutions in the archive will be regarded as the final solutions.
\end{itemize} 
\par By solving (\ref{Forest_monitoring_problem}) with IMOGWO, several benefits are achieved for UAV-enabled IoT in forest monitoring. It effectively addresses the challenges of (\ref{Forest_monitoring_problem}) and improves the algorithm performance. Moreover, IMOGWO also scales efficiently with increasing UAVs and SNs, and its centralized design makes it highly applicable to real-world scenarios, offering a reliable solution for forest monitoring.

\subsubsection{Algorithm analysis} We analyze IMOGWO in terms of complexity and drawbacks as follows.
\par \textit{(a) Complexity:} The computational complexity of the proposed IMOGWO is mainly decided by calculating objective functions and ranking solutions in the archive \cite{zheng2024reliable}. On the one hand, the computational complexity of calculating objective functions is $\mathcal{O}(N_o {Pop})$, where $N_o= 3$ is the number of objective functions in this work. On the other hand, the computational complexity of ranking solutions in the archive is $\mathcal{O}(N_o|\mathcal{A}_{\max}|^2)$, where $|\mathcal{A}_{\max}|$ is the maximum number of solutions in the archive. However, regarding the computational complexity of calculating each objective function as unit is not applicable to solve the formulated optimization framework, whereas we consider the solution dimension as the computational complexity of each objective function. By setting $|\mathcal{A}_{\max}| = Pop$ in this work, the computational complexity of the conventional MOGWO is $\mathcal{O}((6M+6K)G_{\max}{Pop}^2)$. Since the computational complexity of diffusion model updating mechanism, QBL strategy, and discrete solution updating mechanism increases $\mathcal{O}((6M+6K)G_{\max}{Pop}^2)$, respectively, the ultimate computational complexity of IMOGWO is $\mathcal{O}(4(6M+6K)G_{\max}{Pop}^2)$. As can be seen, the computational complexity increases linearly. When $Pop$ is sufficiently large, the computational complexity of the IMOGWO is $\mathcal{O}((6M+6K)G_{\max}{Pop}^2)$, which is the same as the conventional MOGWO.

\par \textit{(b) Drawbacks:} Although IMOGWO is suitable for solving (\ref{Forest_monitoring_problem}), it is still based on the framework of multi-objective heuristic algorithms. Thus, it has the common drawbacks of other multi-objective heuristic algorithms as follows. First, they cannot always guarantee finding the global optimal solution in complex search spaces. Second, their performance can be somewhat reliant on the initial population, often requiring enhancements through the integration of heuristic techniques. Finally, these algorithms are challenging to analyze theoretically, and their parameters can affect the outcomes. It is nearly impossible to claim that any one algorithm is the best for a specific type of problem. However, these methods remain effective for solving NP-hard and non-convex problems, making them practical for real-world applications. In practical scenarios, large-scale optimization problems often arise, where finding the global optimal solution would take an impractical amount of time. Therefore, a viable approach is to use these algorithms, which can provide acceptable solutions in polynomial time while handling multiple constraints.

\section{Simulation Results}
\label{Simulation}
\subsection{Simulation setups}
\par The proposed IMOGWO is conducted by using Matlab to evaluate the performance for coping with (\ref{Forest_monitoring_problem}). For the network setting, we consider two different scales of networks with $6$ and $8$ UAVs, which have $50$ and $100$ SNs, respectively, and these two scales can be recorded as the small-scale network and the large-scale network. The area is assumed as $800$ m $\times$ $800$ m, and SNs are randomly distributed in the area, since the simple and generalized random distribution can reflect various deployment scenarios in practice, which can further enhance the portability and applicability of the scenario. We assume that the trees in the forest are evenly distributed in the forest, and $D_{\text{free}}=4D_{\text{forest}}$. Moreover, the flight altitude range of UAV ($Z_{\min}\sim Z_{\max}$) is $10$ m $\sim 30$ m. Assume that the UAVs fly vertically first and then horizontally, and the climbing velocity, descending velocity, and horizontal velocity are set as $6$ m/s, $2$ m/s, and $10$ m/s, respectively. The minimum and maximum transmission power of SNs are set as $0.1$ W and $1$ W, respectively. The minimum and maximum computing resource of a UAV allocated to an SN are set as $0.5$ GHz and $1$ GHz, respectively, while the local CPU-cycle frequency is $0.1$ GHz. The total transmission power limit of all SNs is $Kp_{\max}/2$, and the safe distance is $5$ m. In addition, the transmission bandwidth and frequency are $1$ MHz and $920$ MHz \cite{myagmardulam2021path}, and the AWGN is $-100$ dBm. The size of computation tasks and computational complexity are set as [$Ca_j=2^{20}w_{1, j}$] bits and [$C_j=100w_{2, j}$] cycles/bits \cite{DBLP:journals/sj/MaoH021, DBLP:journals/tcom/DuZFC18, DBLP:journals/twc/HuWY18, DBLP:journals/twc/YouHCK17}, respectively, where $w_{1, j}=\text{randi}(1, 4)$ and $w_{2, j}=\text{randi}(1, 3)$, in which $\text{randi}(a, b)$ is to generate a random integer between $[a, b]$. Such a random generation reflects the inherent heterogeneity and uncertainty in task demands across the IoT network. For the algorithm setting, $G_{\max}$ and $Pop$ are set as $200$ and $20$, respectively. $\sigma_1$ and $\sigma_2$ are set as $0.1$ and $0.5$, respectively. The penalty constant $C^{\prime}$ is set as $5$. Other related parameters can be found in \cite{8663615}. 
\par Then, several benchmarks are adopted for making comparisons with the proposed IMOGWO including four types, which are random deployment (RD), uniform deployment (UD), multi-objective evolutionary algorithms, and multi-objective swarm intelligence approaches. The details are as follows.

\begin{itemize}
	\item \textbf{RD}: All decision variables are randomly generated within the domain.
	\item \textbf{UD}: UAVs are deployed uniformly in the area, the transmission power is allocated uniformly due to the total transmission power constraint, and the other continuous decision variables are set to the midpoint of their respective bounds, while the discrete decision variables are randomly generated.
	\item \textbf{Multi-objective evolutionary algorithms}: Non-dominated sorting algorithm-III (NSGA-III) \cite{DBLP:journals/tec/DebJ14} and multi-objective evolutionary algorithm based on decomposition (MOEA/D) \cite{DBLP:journals/tec/ZhangL07} are employed as benchmarks, where the random discrete solution updating method in \cite{DBLP:journals/iotj/LiuWSL22} is adopted to update the discrete solution.
	\item \textbf{Multi-objective swarm intelligence approaches}: Multi-objective dragonfly algorithm (MODA) \cite{DBLP:journals/nca/Mirjalili16}, multi-objective salp swarm algorithm (MSSA) \cite{DBLP:journals/aes/MirjaliliGMSFM17}, and the conventional MOGWO \cite{mirjalili2016multi} are also employed as benchmarks, where the random discrete solution updating method is still adopted \cite{DBLP:journals/iotj/LiuWSL22}.
\end{itemize}

\subsection{Optimization results}
\subsubsection{Visualized optimization results}

\par
\begin{figure}[htbp]
	{\includegraphics[width=3.5in]{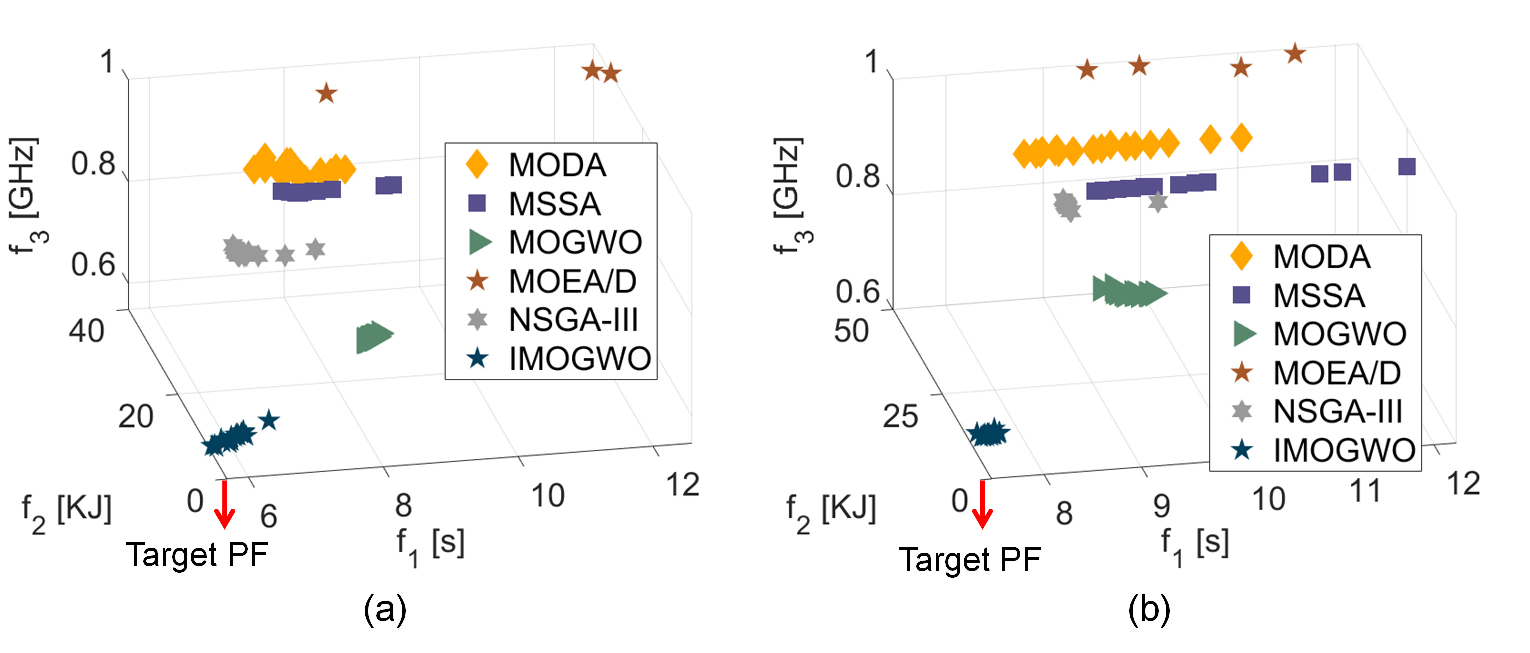}}
	\caption{Solution distribution obtained by different algorithms. (a) Small-scale network. (b) Large-scale network.}
	\label{Paretofig}
\end{figure}

 Figs. \ref{Paretofig}(a) and \ref{Paretofig}(b) show the distribution of the solutions obtained by IMOGWO and other algorithms for the small-scale network and large-scale network, respectively. Obviously, the obtained PF by IMOGWO is much closer to the target PF in both small- and large-scale networks, which represents that it is more suitable for solving (\ref{Forest_monitoring_problem}). As shown in the figures, IMOGWO obtains a better PF compared to the multi-objective evolutionary algorithms and multi-objective swarm intelligence approaches, which means that the obtained PF of IMOGWO can better allow the decision-maker to make trade-offs among different optimization objectives, depending on the urgency and resource availability in the field. For example, in the forest fire monitoring scenario, timeliness is the primary consideration, and hence the decision-maker will select the solution with the minimum $f_1$ from the Pareto optimal solutions, rather than re-running the algorithm. Instead, when the scenario switches to wildlife tracking and behavior monitoring, the requirement for timeliness is reduced, and the economic efficiency of the system becomes the primary concern, which is reflected in the objective functions $f_2$ and $f_3$.

\par
\begin{figure}[tbp]
	\centering
	{\includegraphics[width=3.5in]{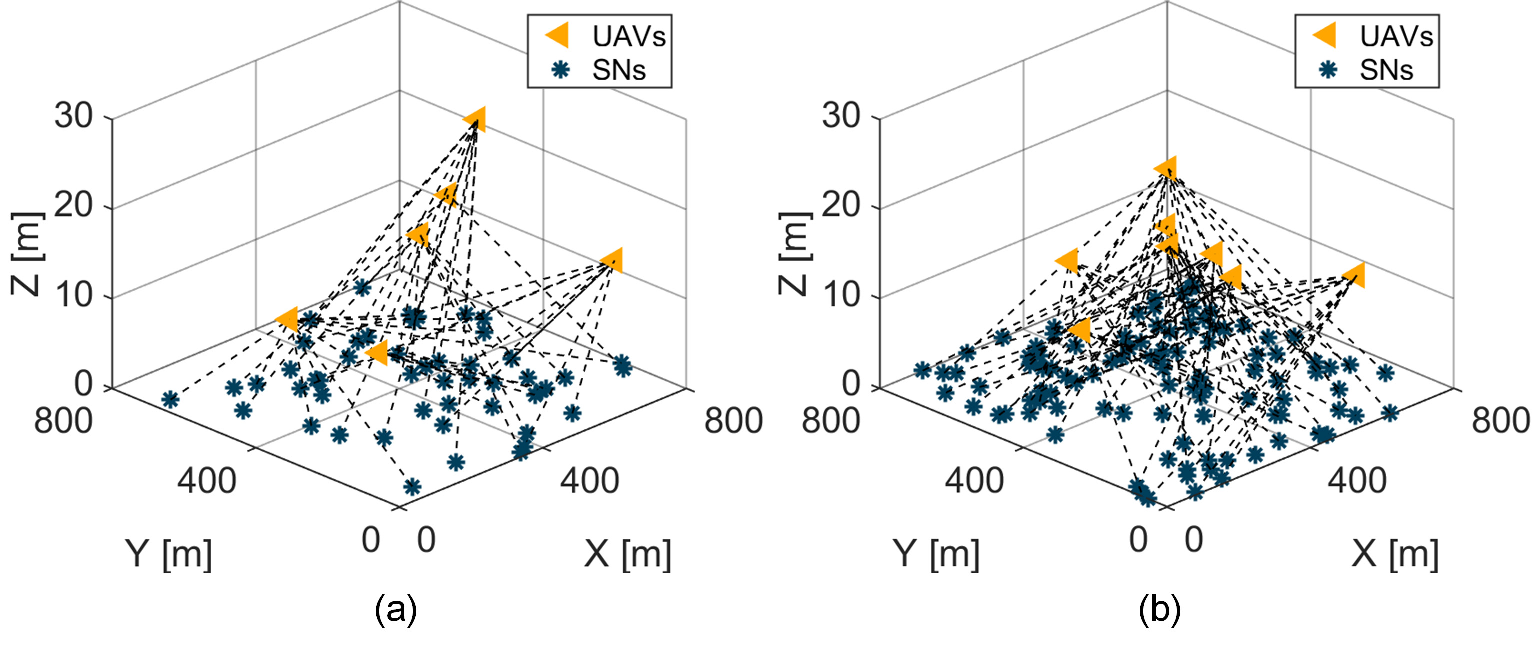}}
	\caption{Optimization results of UAV deployment and UAV-SN offloading assignments. (a) Small-scale network. (b) Large-scale network.}
	\label{Distribution}
\end{figure}

\begin{figure}[htbp]
	\centering
	{\includegraphics[width=3.5in]{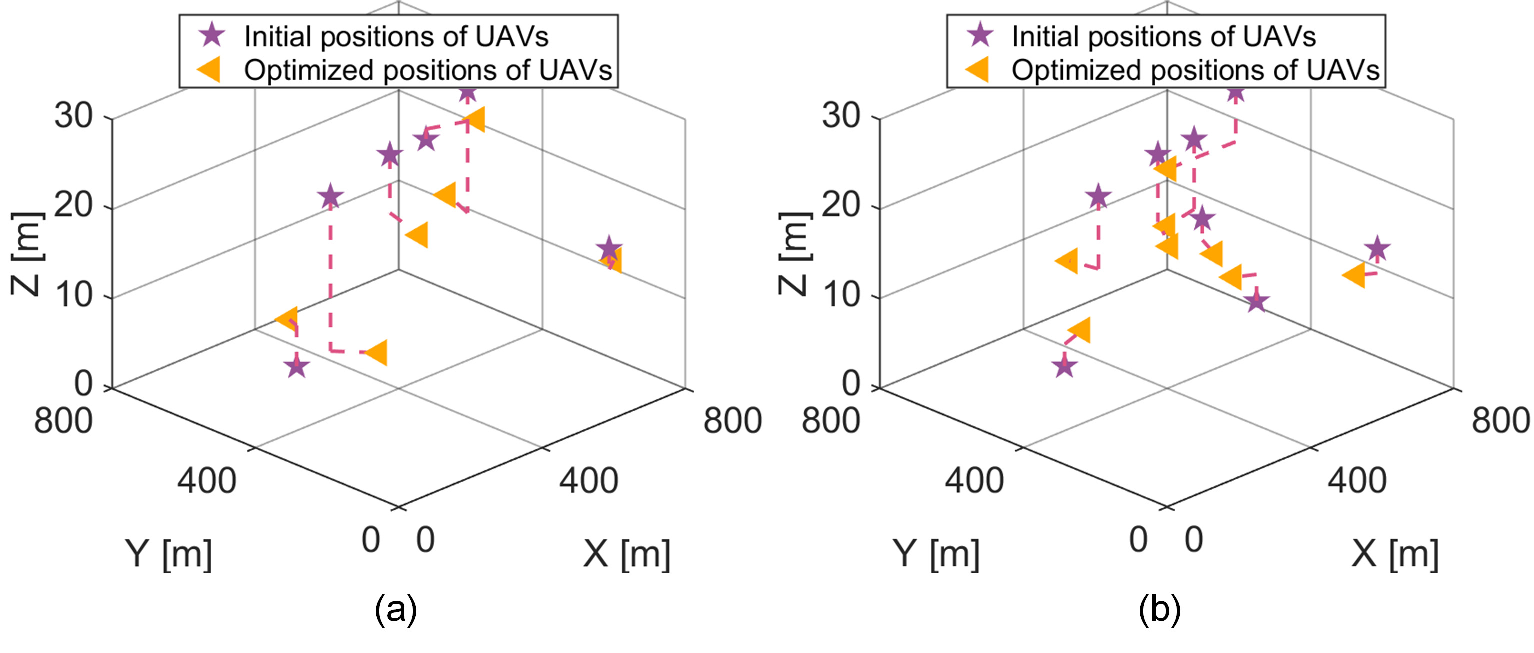}}
	\caption{Optimization results of moving trajectories of UAVs.  (a) Small-scale network. (b) Large-scale network.}
	\label{Trajectory}
\end{figure}

 Fig. \ref{Distribution} shows the optimization results of UAV deployment as well as UAV-SN offloading assignment. It is found that the UAVs tend to gather in areas with high-density SNs, and the number of SNs served by each UAV is similar, which will not increase the communication load of the UAVs. Moreover, Fig. \ref{Trajectory} shows the optimization results of UAV trajectories. As can be seen, in IMOGWO, each UAV strives to reduce the computing delay with a smaller motion energy consumption.


\subsubsection{Stability verification}

\begin{table}[htp]
	\begin{center}
		\caption{Numerical statistical results obtained by different benchmarks in small-scale network}
		\scriptsize
		\setlength{\tabcolsep}{0.2mm}
		{\begin{tabular}{|c|c|c|c|c|c|c|}\hline	
				&Benchmarks &{Mean} &{Std.} &{Max} &{Min} &{$p$-value/gain}\\
				\hline
				\multirow{7}{*}{\textbf {$f_1$ [s]}}
				&\textbf {RD}  		&51.44 &14.84 &62.39 &9.39 &$3.01\rm E-11(+)$\\
				&\textbf {UD}  		&\textbf{6.59} &0.45 &7.86 &6.29 &$6.81\rm E-01(=)$\\
				&\textbf{MODA}  	&8.43 &1.06 &11.38 &6.59 &$1.17\rm E-09(+)$\\
				&\textbf{MSSA}  	&7.53 &0.55 &8.88 &6.59 &$3.64\rm E-08(+)$\\
				&\textbf{MOGWO} 	&8.01 &0.70 &9.30 &6.70 &$9.75\rm E-10(+)$\\
				&\textbf{MOEA/D} 	&11.04 &1.41 &12.57 &8.26 &$3.01\rm E-11(+)$\\
				&\textbf{NSGA-III} 	&10.11 &8.08 &32.49 &\textbf{5.82} &$1.63\rm E-12(+)$\\
				&\textbf{IMOGWO} 	&\textbf{6.60} &\textbf{0.40} &\textbf{7.44} &5.89 & $0.00\%$\\
				\hline
				\multirow{7}{*}{\textbf {$f_2$ [KJ]}}
				&\textbf {RD}  		&162.12 &50.03 &228.95 &29.92 &$3.01\rm E-11(+)$\\
				&\textbf {UD}  		&38.82 &\textbf{0.00} &38.82 &38.82 &$1.21\rm E-12(+)$\\
				&\textbf{MODA}  	&26.96 &4.19 &34.68 &17.35 &$3.01\rm E-11(+)$\\
				&\textbf{MSSA}  	&17.16 &3.94 &26.19 &8.83 &$1.32\rm E-10(+)$\\
				&\textbf{MOGWO} 	&17.55 &4.88 &26.41 &7.29 &$1.41\rm E-09(+)$\\
				&\textbf{MOEA/D} 	&35.97 &6.66 &48.56 &22.44 &$3.01\rm E-11(+)$\\
				&\textbf{NSGA-III} 	&19.41 &14.46 &68.88 &9.24 &$4.57\rm E-09(+)$\\
				&\textbf{IMOGWO} 	&\textbf{8.01} &2.28 &\textbf{13.02} &\textbf{4.04} &$53.32\%$\\
				\hline
				\multirow{7}{*}{\textbf {$f_3$ [GHz]}}				
				&\textbf {RD}  		&4.54 &1.20 &4.99 &0.99 &$3.01\rm E-11(+)$\\
				&\textbf {UD}  		&0.75 &\textbf{0.00} &0.75 &0.75 &$1.21\rm E-12(+)$\\
				&\textbf{MODA}  	&0.94 &0.02 &0.98 &0.90 &$3.01\rm E-11(+)$\\
				&\textbf{MSSA}  	&0.90 &0.02 &0.94 &0.84 &$3.01\rm E-11(+)$\\
				&\textbf{MOGWO} 	&0.61 &0.03 &0.69 &0.55 &$4.99\rm E-09(+)$\\
				&\textbf{MOEA/D} 	&0.93 &0.03 &0.98 &0.86 &$3.01\rm E-11(+)$\\
				&\textbf{NSGA-III} 	&1.32 &1.15 &4.31 &0.82 &$3.01\rm E-11(+)$\\
				&\textbf{IMOGWO} 	&\textbf{0.55} &0.02 &\textbf{0.62} &\textbf{0.53} &$9.83\%$\\
				\hline
		\end{tabular}}
		\label{Num_small}
	\end{center}
\end{table}

\begin{table}[htbp]
	\begin{center}
		\caption{Numerical statistical results obtained by different benchmarks in large-scale network}
		\scriptsize
		\setlength{\tabcolsep}{0.2mm}
		{\begin{tabular}{|c|c|c|c|c|c|c|}\hline	
				&Benchmarks &{Mean} &{Std.} &{Max} &{Min} &{$p$-value/gain}\\
				\hline
				\multirow{7}{*}{\textbf {$f_1$ [s]}}
				&\textbf {RD}  		&57.39 &10.19 &82.17 &12.18 &$3.01\rm E-11(+)$\\
				&\textbf {UD}  		&\textbf{7.97} &0.92 &9.88 &\textbf{6.34} &$4.50\rm E-01(=)$\\
				&\textbf{MODA}  	&11.85 &9.39 &46.66 &7.89 &$4.61\rm E-10(+)$\\
				&\textbf{MSSA}  	&8.95 &0.55 &9.94 &7.88  &$1.69\rm E-09(+)$\\
				&\textbf{MOGWO} 	&9.13 &0.38 &9.61 &7.87 &$1.61\rm E-10(+)$\\
				&\textbf{MOEA/D} 	&11.44 &1.14 &12.58 &8.15 &$7.34\rm E-11(+)$\\
				&\textbf{NSGA-III} 	&25.39 &13.73 &39.34 &7.57 &$5.57\rm E-10(+)$\\
				&\textbf{IMOGWO} 	&\textbf{7.88} &\textbf{0.34} &\textbf{8.54} &7.05 &$0.00\%$\\
				\hline
				\multirow{7}{*}{\textbf {$f_2$ [KJ]}}				
				&\textbf {RD}  		&211.42 &43.19 &281.82 &42.03 &$3.01\rm E-11(+)$\\
				&\textbf {UD}  		&43.86 &\textbf{0.00} &43.86 &43.86 &$1.21\rm E-12(+)$\\
				&\textbf{MODA}  	&42.54 &23.08 &126.44 &22.70 &$3.01\rm E-11(+)$\\
				&\textbf{MSSA}  	&25.23 &4.89 &35.57 &14.34 &$7.38\rm E-10(+)$\\
				&\textbf{MOGWO} 	&28.04 &5.35 &41.50 &16.33 &$7.38\rm E-11(+)$\\
				&\textbf{MOEA/D} 	&43.81 &7.61 &57.68 &28.32 &$3.01\rm E-11(+)$\\
				&\textbf{NSGA-III} 	&57.67 &31.57 &118.86 &19.41 &$8.99\rm E-11(+)$\\
				&\textbf{IMOGWO} 	&\textbf{14.68} &3.02 &\textbf{20.21} &\textbf{9.17} &$41.81\%$\\
				\hline
				\multirow{7}{*}{\textbf {$f_3$ [GHz]}}				
				&\textbf {RD}  		&4.84 &0.72 &4.99 &0.99 &$3.01\rm E-11(+)$\\
				&\textbf {UD}  		&0.75 &\textbf{0.00} &0.75 &0.75 &$1.21\rm E-12(+)$\\
				&\textbf{MODA}  	&1.22 &0.97 &4.81 &0.92 &$3.01\rm E-11(+)$\\
				&\textbf{MSSA}  	&0.93 &0.01 &0.96 &0.89 &$3.01\rm E-11(+)$\\
				&\textbf{MOGWO} 	&0.63 &0.02 &0.70 &0.59 &$6.52\rm E-08(+)$\\
				&\textbf{MOEA/D} 	&0.97 &0.01 &1.00 &0.91 &$3.01\rm E-11(+)$\\
				&\textbf{NSGA-III} 	&2.96 &1.68 &4.49 &0.89 &$3.01\rm E-11(+)$\\
				&\textbf{IMOGWO} 	&\textbf{0.58} &0.02 &\textbf{0.62} &\textbf{0.53} &$7.93\%$\\
				\hline
		\end{tabular}}
		\label{Num_large}
	\end{center}
\end{table}

\par Due to the random nature of multi-objective heuristic algorithms \cite{DBLP:journals/isci/ZhaoWJC22}, we run IMOGWO and benchmarks $30$ times independently and repeatedly for avoiding the random bias\footnote{In IMOGWO, each iteration is a sampling process, like generating random UAV positions. A sample size of $30$ is chosen based on the central limit theorem, balancing accuracy and simulation efficiency.}, and then the Wilcoxon rank sum test is used to determine whether significant differences are present or absent among these algorithms \cite{DBLP:journals/tcyb/ZhanLC0CS13}. Under the random distribution of UAV initial positions, the performance of these benchmarks is shown in Tables \ref{Num_small} and \ref{Num_large}, in which ``Mean'', ``Std.'', ``Max'' and ``Min'' are the mean value, standard deviation, maximum value, and minimum value. In addition, a $p$-value of less than $0.05$ indicates the presence of a significant difference between IMOGWO and other benchmarks on the corresponding optimization objective. Otherwise, it indicates that there is no significant difference between IMOGWO and other benchmarks in the corresponding optimization objective. Moreover, the symbols ``$+$", ``$=$", and ``$-$" indicate that IMOGWO significantly outperforms other benchmarks, IMOGWO has no significant difference from other benchmarks, and other benchmarks significantly outperform IMOGWO, respectively. ``gain'' means that the performance improvement of IMOGWO compared to the second best value obtained by other benchmarks on the corresponding optimization objective.

\par As can be seen, for a small-scale network, compared to the suboptimal benchmark, IMOGWO reduces the motion energy consumption and the computing resource by $53.32\%$ and $9.83\%$, respectively, while maintaining computing delay at the same level. Moreover, for a large-scale network, IMOGWO achieves reductions of $41.81\%$ in motion energy consumption and $7.93\%$ in computing resource, with the computing delay also remaining comparable. It is worth noting that some of the maximum values reported in the tables exceed their upper bounds. For example, the maximum value of $f_3$ obtained by NSGA-III in the small-scale network exceeds its upper bound. This is because the final solutions obtained by NSGA-III still fail to satisfy constraints $\mathcal{C}_9$ or $\mathcal{C}_{10}$. In contrast, IMOGWO satisfies $\mathcal{C}_9$ and $\mathcal{C}_{10}$ in all $30$ independent and repeated tests, demonstrating its stability.

\subsubsection{Ablation verification}
\par To evaluate the effectiveness of each improved factor, we conduct the ablation simulations. Specifically, we utilize MOGWO, improved MOGWO with a diffusion model updating mechanism (MOGWO-1), improved MOGWO with a QBL mechanism (MOGWO-2), and improved MOGWO with a discrete solution processing mechanism (MOGWO-3) to solve the formulated framework, and the $30$ independent tests are also conducted to compare the optimization results. As shown in Tables \ref{Ablation1} and \ref{Ablation2}, MOGWO-1 can improve the results of all optimization objectives, while MOGWO-2 and MOGWO-3 can improve the results of part of the optimization objectives due to the trade-offs between the optimization objectives. However, IMOGWO, combining all the advantages of MOGWO-1, MOGWO-2, and MOGWO-3, can further improve their results, and hence IMOGWO is effective and all improved mechanisms are necessary.

\begin{table}[htbp]
	\begin{center}
		\caption{Ablation simulation statistical results in small-scale network}
		\scriptsize
		\setlength{\tabcolsep}{0.2mm}
		{\begin{tabular}{|c|c|c|c|c|c|}\hline	
				&Benchmarks &{Mean} &{Std.} &{Max} &{Min} \\
				\hline
				\multirow{5}{*}{\textbf {$f_1$ [s]}}
				&\textbf{MOGWO} 	&8.01 &0.70 &9.30 &6.70  \\
				&\textbf{MOGWO-1} 	&7.09 &\textbf{0.37} &7.83 &6.50  \\
				&\textbf{MOGWO-2} 	&7.79 &0.68 &9.48 &6.52  \\
				&\textbf{MOGWO-3} 	&7.96 &0.48 &8.95 &7.00  \\
				&\textbf{IMOGWO} 	&\textbf{6.60} &0.40 &\textbf{7.44} &\textbf{5.89} \\
				\hline
				\multirow{5}{*}{\textbf {$f_2$ [KJ]}}
				&\textbf{MOGWO} 	&17.55 &4.88 &26.41 &7.29  \\   
				&\textbf{MOGWO-1} 	&8.17 &\textbf{1.69} &\textbf{11.28} &5.56  \\
				&\textbf{MOGWO-2} 	&14.84 &4.11 &23.09 &8.28  \\
				&\textbf{MOGWO-3} 	&16.89 &4.10 &27.98 &10.32  \\
				&\textbf{IMOGWO} 	&\textbf{8.01} &2.28 &13.02 &\textbf{4.04}  \\
				\hline
				\multirow{5}{*}{\textbf {$f_3$ [GHz]}}				
				&\textbf{MOGWO} 	&0.61 &0.03 &0.69 &0.55 \\
				&\textbf{MOGWO-1} 	&\textbf{0.55} &\textbf{0.01} &\textbf{0.57} &\textbf{0.52} \\
				&\textbf{MOGWO-2} 	&0.61 &0.04 &0.73 &0.54  \\
				&\textbf{MOGWO-3} 	&0.60 &0.02 &0.69 &0.55  \\
				&\textbf{IMOGWO} 	&\textbf{0.55} &\textbf{0.01} &\textbf{0.57} &\textbf{0.52}  \\
				\hline
		\end{tabular}}
		\label{Ablation1}
	\end{center}
\end{table}

\begin{table}[htbp]
	\begin{center}
		\caption{Ablation simulation statistical results in large-scale network}
		\scriptsize
		\setlength{\tabcolsep}{0.2mm}
		{\begin{tabular}{|c|c|c|c|c|c|}\hline	
				&Benchmarks &{Mean} &{Std.} &{Max} &{Min} \\
				\hline
				\multirow{5}{*}{\textbf {$f_1$ [s]}}
				&\textbf{MOGWO} 	&9.13 &0.38 &9.61 &7.87  \\
				&\textbf{MOGWO-1} 	&8.13 &\textbf{0.23} &8.65 &7.68  \\
				&\textbf{MOGWO-2} 	&8.92 &0.41 &9.84 &7.85  \\
				&\textbf{MOGWO-3} 	&9.01 &0.58 &10.09 &7.89  \\
				&\textbf{IMOGWO} 	&\textbf{7.88} &0.34 &\textbf{8.54} &\textbf{7.05} \\
				\hline
				\multirow{5}{*}{\textbf {$f_2$ [KJ]}}
				&\textbf{MOGWO} 	&28.04 &5.35 &41.50 &16.33 \\
				&\textbf{MOGWO-1} 	&15.31 &3.97 &22.14 &\textbf{8.96}  \\
				&\textbf{MOGWO-2} 	&24.92 &4.15 &33.22 &17.72  \\
				&\textbf{MOGWO-3} 	&26.50 &6.13 &42.04 &12.71  \\
				&\textbf{IMOGWO} 	&\textbf{14.68} &\textbf{3.02} &\textbf{20.21} &9.17 \\
				\hline
				\multirow{5}{*}{\textbf {$f_3$ [GHz]}}				
				&\textbf{MOGWO} 	&0.63 &0.02 &0.70 &0.59  \\
				&\textbf{MOGWO-1} 	&\textbf{0.58} &\textbf{0.01} &0.63 &0.56  \\
				&\textbf{MOGWO-2} 	&0.66 &0.03 &0.75 &0.60  \\
				&\textbf{MOGWO-3} 	&0.65 &0.03 &0.79 &0.58  \\
				&\textbf{IMOGWO} 	&\textbf{0.58} &0.02 &\textbf{0.62} &\textbf{0.53} \\
				\hline
		\end{tabular}}
		\label{Ablation2}
	\end{center}
\end{table}
\subsubsection{Convergence verification}
\par It is worth noting that proving convergence is challenging due to the stochastic nature of the multi-objective swarm intelligence approaches \cite{DBLP:journals/tec/DebJ14}, such as IMOGWO. Moreover, it is difficult to give a direct convergence curve for multiple optimization objectives \cite{10812989}. Hence, the solution distribution, inverted generational distance (IGD), and variation of objective functions are adopted to evaluate the convergence of IMOGWO \cite{10812989}.

\par
\begin{figure}[htbp]
	\centering
	{\includegraphics[width=3.5in]{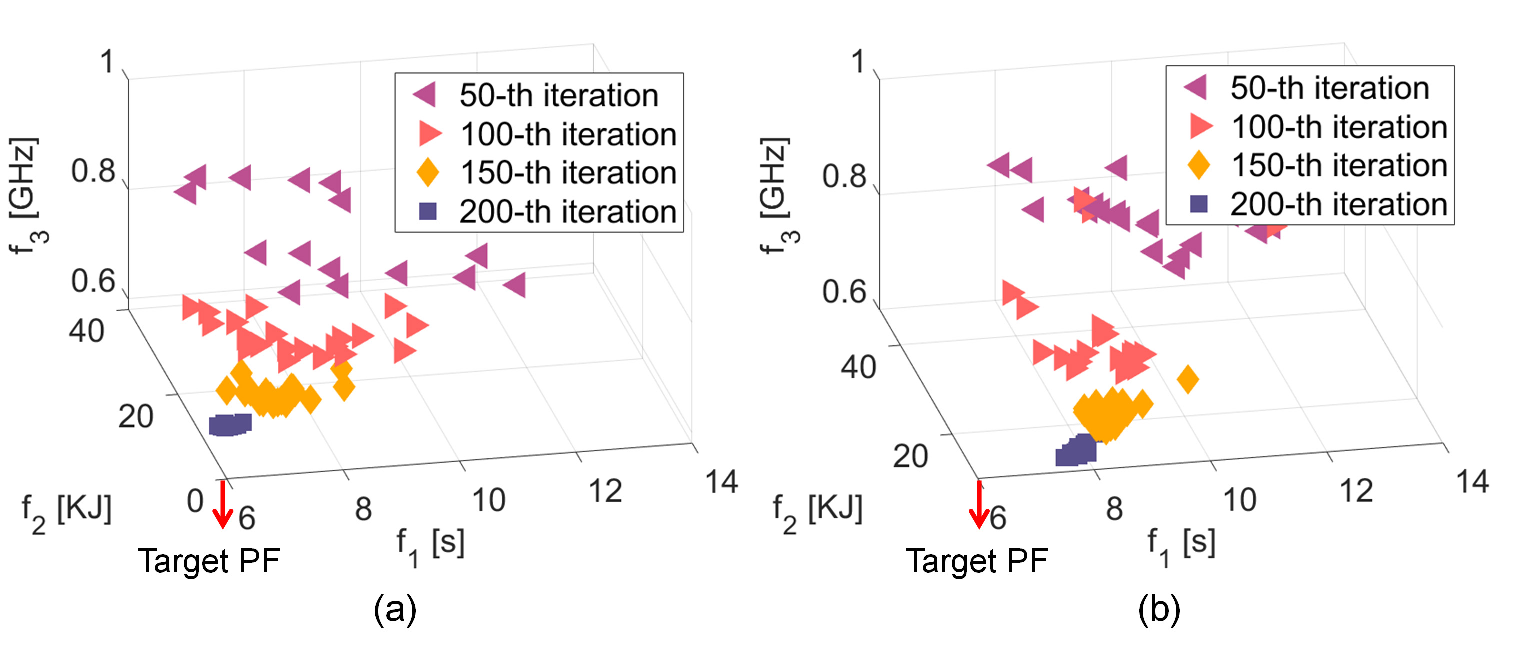}}
	\caption{Solution distribution obtained by IMOGWO. (a) Small-scale network. (b) Large-scale network.}
	\label{PF_advancement_process}
\end{figure}

 \textit{(a) Solution distribution:} The formulated optimization framework with the three optimization objectives is a multi-objective optimization problem with trade-offs, making it challenging to achieve optimal values for all objectives simultaneously. Thus, utilizing the solution distribution during the iteration is a logical approach to validate the convergence of the algorithm. For this purpose, Figs. \ref{PF_advancement_process}(a) and \ref{PF_advancement_process}(b) show the advanced process of PF obtained by IMOGWO, which means that IMOGWO searches iteratively for the target PF. As can be seen, the PF obtained at the $200$-th iteration surpasses that obtained at the $150$-th iteration in both scales of networks, indicating an improvement in solution quality with more iterations, and the improvement gradually decreases, which means that IMOGWO converges.

\par
\begin{figure}[htbp]
	\centering
	{\includegraphics[width=3.5in]{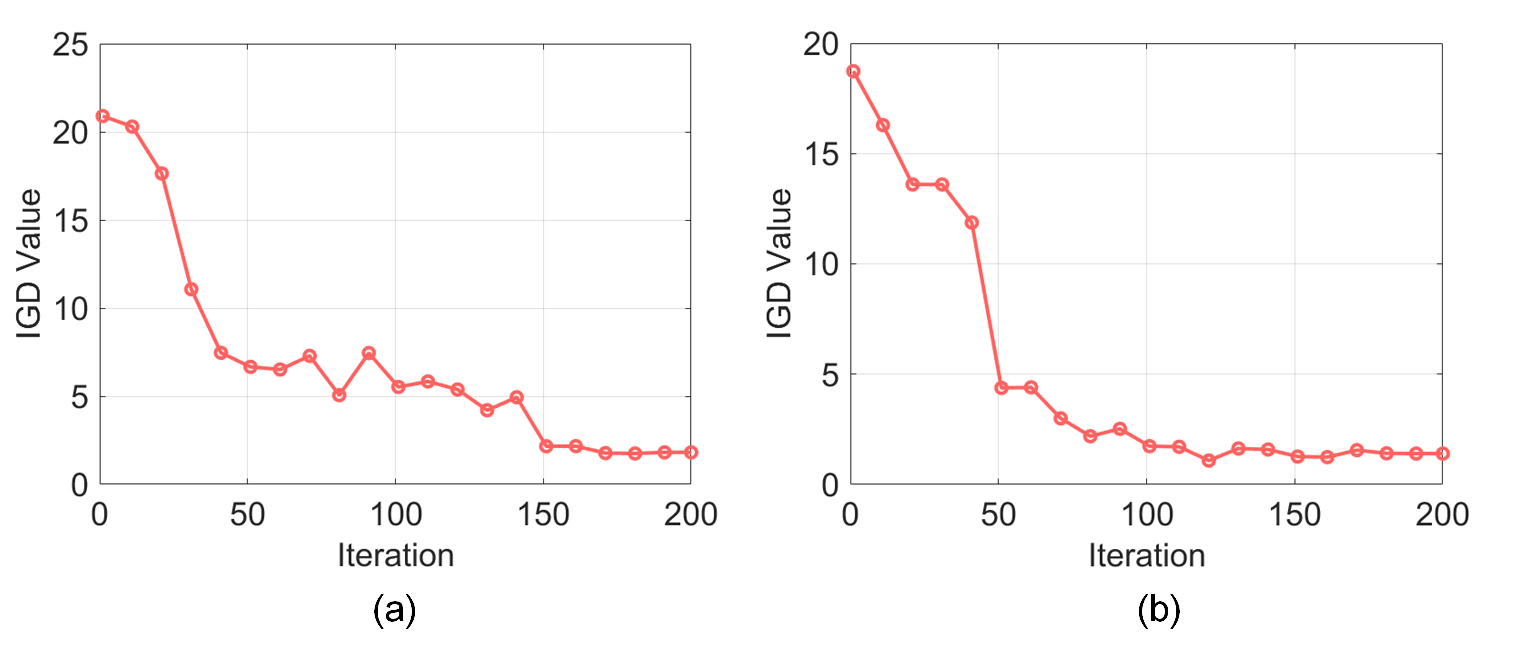}}
	\caption{IGD curve. (a) Small-scale network. (b) Large-scale network.}
	\label{IGD}
\end{figure}

 \textit{(b) IGD:} IGD represents the average distance between the obtained solution set and the true PF. Since the true PF is difficult to obtain, we use the non-dominated solutions from multiple simulations to construct an approximate PF. Figs. \ref{IGD}(a) and \ref{IGD}(b) show the IGD curves, where the IGD values stabilize for both scales of the networks, indicating that IMOGWO converges effectively.
\par
\begin{figure}[htbp]
	\centering
	{\includegraphics[width=3.5in]{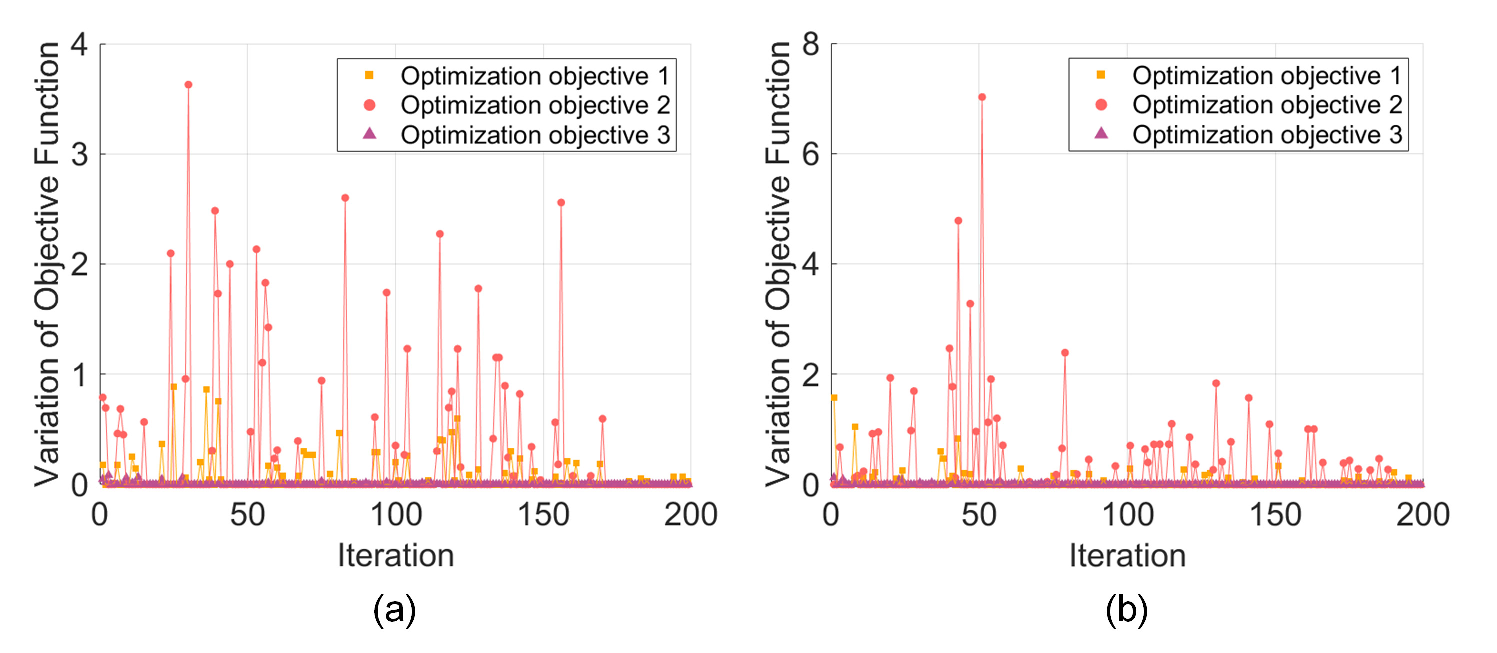}}
	\caption{Variation of objective functions. (a) Small-scale network. (b) Large-scale network.}
	\label{Variation_of_objective_functions}
\end{figure}

 \textit{(c) Variation of objective functions:} The variation of objective functions can reflect convergence performance \cite{10812989}. Specifically, we calculate the variation of objective functions at each iteration by tracking the minimum objective value within the Pareto set. Figs. \ref{Variation_of_objective_functions}(a) and \ref{Variation_of_objective_functions}(b) illustrate that the variations in all three objective functions progressively diminish as iteration proceeds, suggesting that the algorithm is approaching convergence.

\subsubsection{Robustness verification} In this part, the impacts of different initial UAV distributions and UAV drifts are verified.
\par \textit{(a) Impact of different initial UAV distributions:} To assess the robustness of the algorithm, we further evaluate its performance under a Gaussian distribution (\_G), and an exponential distribution (\_E). The corresponding numerical results for small- and large-scale networks are shown in Table \ref{table: different_initial_distributions}. Similarly, we conduct $30$ independent and repeated tests, and then the Wilcoxon rank sum test is also used to determine whether significant differences are present or absent among these algorithms \cite{DBLP:journals/tcyb/ZhanLC0CS13}. As shown in the table, for the small-scale network, IMOGWO can obtain the best performance on $f_1$ and $f_3$, while the gap between IMOGWO and the optimal value on $f_2$ is not large, when the initial distribution of UAVs is \_G. However, when the initial distribution of UAVs is \_E, IMOGWO can obtain the best performance on the three optimization objectives. Moreover, for the large-scale network, IMOGWO can also obtain the best performance on the three optimization objectives under \_G and \_E, which means that IMOGWO is more suitable to solve the formulated framework, especially in the large-scale network. Thus, IMOGWO achieves the best overall performance under \_G and \_E, demonstrating strong adaptability to diverse initial conditions. This robustness is crucial in forest fire monitoring, where UAVs may launch from varied positions due to terrain or emergency needs, ensuring reliable and timely decisions.
\begin{table}[htbp]
	\begin{center}
		\caption{Numerical statistical results for different initial distributions of UAVs}
		\tiny
		\setlength{\tabcolsep}{0.1mm}
		{\begin{tabular}{|c|c|c|c|c|c|c|}\hline					
				\textbf {Algorithm} &{\textbf {$f_1$ [s]}} &$p$-value &{\textbf {$f_2$ [KJ]}}&$p$-value &{\textbf {$f_3$ [GHz]}}&$p$-value \\
				\hline
				&\multicolumn{6}{c|}{\textbf{Small-scale network}} \\\hline	
				\textbf{MODA\_G}  	&8.16 &$8.35\rm E-08(+)$ &16.27 &$8.99\rm E-11(+)$ &0.94 &$3.01\rm E-11(+)$\\
				\textbf{MSSA\_G}  	&7.51 &$7.08\rm E-08(+)$ &11.33 &$3.83\rm E-06(+)$ &0.89 &$3.01\rm E-11(+)$ \\
				\textbf{MOGWO\_G} 	&8.22 &$4.57\rm E-09(+)$ &16.72 &$2.15\rm E-10(+)$ &0.61 &$4.31\rm E-08(+)$ \\
				\textbf{MOEA/D\_G} 	&11.02 &$3.01\rm E-11(+)$ &26.63 &$3.01\rm E-11(+)$ &0.93 &$3.01\rm E-11(+)$  \\
				\textbf{NSGA-III\_G}&8.40 &$1.29\rm E-01(+)$ &\textbf{7.23} &$3.98\rm E-04(-)$ &1.08 &$3.01\rm E-11(+)$  \\
				\textbf{IMOGWO\_G} 	&\textbf{6.64} & &8.25 & &\textbf{0.56} & \\
				\hline
				\textbf{MODA\_E}  	&8.29 &$3.32\rm E-06(+)$ &30.36 &$3.01\rm E-11(+)$ &0.94 &$3.01\rm E-11(+)$ \\
				\textbf{MSSA\_E}  	&7.61 &$7.19\rm E-05(+)$ &20.06 &$3.01\rm E-11(+)$ &0.91 &$3.01\rm E-11(+)$  \\
				\textbf{MOGWO\_E} 	&7.59 &$1.40\rm E-04(+)$ &5.32 &$4.68\rm E-08(+)$ &\textbf{0.60} &$2.97\rm E-01(=)$ \\
				\textbf{MOEA/D\_E} 	&12.44 &$3.01\rm E-11(+)$ &44.68 &$3.01\rm E-11(+)$ &1.05 &$3.01\rm E-11(+)$ \\
				\textbf{NSGA-III\_E} 	&\textbf{9.74} &$9.10\rm E-01(=)$ &22.96 &$3.01\rm E-11(+)$ &1.30 &$3.01\rm E-11(+)$\\
				\textbf{IMOGWO\_E} 	&\textbf{6.81} & &\textbf{3.41} & &\textbf{0.61} & \\
				\hline		
				& \multicolumn{6}{c|}{\textbf{Large-scale network}} \\\hline	
				\textbf{MODA\_G}  &9.75 &$1.09\rm E-09(+)$ &26.47 &$4.50\rm E-11(+)$ &0.97 &$2.95\rm E-11(+)$  \\
				\textbf{MSSA\_G}  &8.92 &$5.96\rm E-09(+)$ &19.53 &$3.08\rm E-08(+)$ &0.93 &$3.01\rm E-11(+)$ \\
				\textbf{MOGWO\_G} &9.16 &$3.82\rm E-10(+)$ &21.57 &$2.43\rm E-09(+)$ &0.64 &$1.82\rm E-02(+)$ \\
				\textbf{MOEA/D\_G} &11.06 &$3.68\rm E-11(+)$ &35.59 &$3.01\rm E-11(+)$ &0.98 &$2.98\rm E-11(+)$ \\
				\textbf{NSGA-III\_G}&23.92 &$2.31\rm E-06(+)$ &30.83 &$1.40\rm E-03(+)$ &2.83 &$3.01\rm E-11(+)$ \\
				\textbf{IMOGWO\_G} &\textbf{7.95} & &\textbf{12.80} & &\textbf{0.62} & \\
				\hline
				\textbf{MODA\_E}  &10.74 &$2.19\rm E-08(+)$ &46.56 &$3.01\rm E-11(+)$ &1.09 &$3.01\rm E-11(+)$ \\
				\textbf{MSSA\_E}  &9.18 &$1.42\rm E-08(+)$ &29.37 &$3.01\rm E-11(+)$ &0.93 &$3.01\rm E-11(+)$ \\
				\textbf{MOGWO\_E} &8.88 &$7.69\rm E-08(+)$ &11.98 &$8.89\rm E-10(+)$ &\textbf{0.62} &$6.35\rm E-02(=)$ \\
				\textbf{MOEA/D\_E}  &10.98 &$6.05\rm E-11(+)$ &51.48 &$3.01\rm E-11(+)$ &0.98 &$2.80\rm E-11(+)$ \\
				\textbf{NSGA-III\_E} &23.37 &$2.83\rm E-04(+)$ &80.34 &$3.01\rm E-11(+)$ &2.73 &$3.01\rm E-11(+)$ \\
				\textbf{IMOGWO\_E} 	&\textbf{8.08} & &\textbf{7.26} & &\textbf{0.64} & \\
				\hline		
		\end{tabular}}
		\label{table: different_initial_distributions}
	\end{center}
\end{table}

\par \textit{(b) Impact of UAV drifts:} The UAVs may drift from the assigned positions due to the environmental factors, such as the wind, which means that $f_1$ (minimizing the maximum computing delay) and $f_2$ (minimizing the total motion energy consumption) will be impacted. Thus, we evaluate the impact of UAV position drifts by using the following methods. According to \cite{DBLP:journals/tmc/LiSDW24}, we generate some position drifts by using the normal distribution. Note that the maximum values of UAV drifts are set to $0.2$ m, $0.4$ m, $0.6$ m, $0.8$ m and $1$ m, respectively, and the simulations are repeated $30$ times to avoid the random bias. Table \ref{table: UAV_drift} shows the performance gaps between the original and position-drifted conditions, and the results show that there is no difference on $f_1$ and the differences on $f_2$ are also not very obvious. The reasons are as follows: IMOGWO deploys UAVs near optimal service positions. Thus, even with slight drift, UAVs remain in suboptimal yet effective positions for serving SNs. In addition, $f_1$ is defined as the maximum of the local computing delay and the offloading delay, while UAV position drift only affects the offloading delay, which means that $f_1$ may not change if the local computing delay is larger than the offloading delay. Moreover, since the distance of UAV position shift is not large, the change in $f_2$ is also small. Thus, the proposed approach is robust.
\begin{table}[H]
	\begin{center}
		\caption{Numerical statistical results for UAV position drift in small- and large-scale networks}
		\tiny
		\setlength{\tabcolsep}{2mm}
		{\begin{tabular}{|c|c|c|c|c|c|}\hline	
				
				&\multicolumn{2}{c|}{\textbf{Small-scale network}} & \multicolumn{2}{c|}{\textbf{Large-scale network}}\\\hline	
				\textbf {UAV drift distance} &{\textbf {$f_1$ [s]}} &{\textbf {$f_2$ [KJ]}} &{\textbf {$f_1$ [s]}} &{\textbf {$f_2$ [KJ]}} \\
				\hline
				\textbf{0.0 m}  &6.46 &7.99 &7.53 &7.68 \\
				\textbf{0.2 m}  &6.46 &7.99 &7.53 &7.68 \\
				\textbf{0.4 m}  &6.46 &7.99 &7.53 &7.68 \\
				\textbf{0.6 m}  &6.46 &7.99 &7.53 &7.69 \\				
				\textbf{0.8 m}  &6.46 &7.99 &7.53 &7.68 \\
				\textbf{1.0 m}  &6.46 &8.01 &7.53 &7.70 \\
				\hline			
		\end{tabular}}
		\label{table: UAV_drift}
	\end{center}
\end{table}

\subsubsection{Running time and practicality tests}
\par We give the running time of IMOGWO and the comparison algorithms in Table \ref{Time_IMOGWO}. As can be seen, the running time of IMOGWO does not increase largely compared to the conventional MOGWO, and the gaps between IMOGWO and other comparison algorithms are also not very large. In addition, the proposed UAV-based deployment is usually performed off-line. Thus, the running time of IMOGWO is feasible.

\begin{table}[htbp]
	\begin{center}
		\caption{Numerical statistical results of running times}
		\tiny
		\label{Time_IMOGWO}	
		\begin{tabular}{|c|c|c|}\hline
			\textbf{Algorithms}&{\textbf{Small-scale network [s]}}& {\textbf {Large-scale network [s]}}
			\\\hline
			\textbf{MODA} & {6.28} & {10.38}\\
			\textbf{MSSA} & {1.16} & {1.86}\\
			\textbf{MOGWO} & {4.48} & {5.88}\\
			\textbf{MOEA/D} & {1.94} & {2.67} \\
			\textbf{NSGA-III} & {8.43} & {16.96} \\
			\textbf{IMOGWO} & {6.62} & {10.36} \\
			\hline
		\end{tabular}
	\end{center}
\end{table}

\par Moreover, we reimplement the proposed IMOGWO in Python and deploy it on the Raspberry Pi 4B platform, which is regarded as the CPU of UAVs \cite{DBLP:journals/wcl/ZhouHJMD21, pan2025cooperative}. This hardware setup allows us to evaluate the algorithm performance under realistic resource constraints. Furthermore, by setting identical random seeds across different UAVs, the system supports decentralized decision-making, enabling multiple UAVs to independently execute the optimization algorithm while maintaining global consistency. This property facilitates distributed deployment with low communication overhead, which is particularly beneficial in large-scale forest monitoring scenarios. In addition, since calculating the objective functions is usually replaced by the agent model \cite{pan2025cooperative}, we regard it as the unit time and ignore it when executing on the Raspberry Pi 4B. Under these circumstances, the execution times of Raspberry Pi 4B are $99.62$ s and $176.04$ s for the small- and large-scale networks, respectively, and both values fall within feasible ranges. These results demonstrate that IMOGWO can be efficiently executed on the Raspberry Pi platform, which confirms the practical applicability and ensures that its simulation performance can be consistently reproduced in real deployment scenarios, thereby effectively bridging the gap between theoretical design and practical implementation.

\section{Conclusion}
\label{Conclusion}

\par In this paper, a UAV-enabled IoT system is considered to efficiently monitor the forests. To balance the trade-offs between monitoring performance, energy budget, and computing resource, we formulate a multi-objective optimization framework that aims to jointly minimize the maximum computing delay, minimize the total motion energy consumption, and minimize the maximum computing resource. As the formulated framework is a hybrid multi-objective optimization framework with both continuous and discrete solution spaces, we propose an IMOGWO with a diffusion model updating mechanism, a QBL strategy, and a discrete solution updating mechanism to improve the performance of the algorithm. Simulation results demonstrate that IMOGWO reduces the motion energy consumption and the computing resource by $53.32\%$ and $9.83\%$, respectively, while maintaining the computing delay at the same level for a small-scale network with $6$ UAVs and $50$ SNs compared to the suboptimal benchmark. Similarly, for a large-scale network with $8$ UAVs and $100$ SNs, IMOGWO achieves reductions of $41.81\%$ in motion energy consumption and $7.93\%$ in computing resource, with the computing delay also remaining comparable.

\bibliographystyle{ieeetr}
\bibliography{ref-MyUAV}

\end{document}